\documentclass[
reprint,
 amsmath,amssymb,
 aps,
pra,
]{revtex4-1}

\usepackage{graphicx}
\usepackage{dcolumn}
\usepackage{bm}

\DeclareMathAlphabet{\mathpzc}{OT1}{pzc}{m}{it} 
\usepackage{braket,color,comment,amsmath}

\begin{document}

\preprint{APS/123-QED}

\title{A fluctuation theorem for time-series of signal-response models\\ with the backward transfer entropy}

\author{Andrea Auconi$^{1,2}$, Andrea Giansanti$^{2,3}$ and Edda Klipp$^{1,}$}
\email{edda.klipp@rz.hu-berlin.de}

\affiliation{$1$ Theoretische Biophysik, Humboldt-Universit\"at zu Berlin, Invalidenstra\ss e 42, D-10115 Berlin, Germany\\
	$2$ Dipartimento di Fisica, Sapienza Universit\`a di Roma, Rome, Italy\\
	$3$ INFN, Sezione di Roma 1, Rome, Italy}

\date{\today}

\begin{abstract}
The irreversibility of trajectories in stochastic dynamical systems is linked to the structure of their causal representation in terms of Bayesian networks.
We consider stochastic maps resulting from a time discretization with interval $\tau$ of signal-response models, and we find an integral fluctuation theorem that sets the backward transfer entropy as a lower bound to the conditional entropy production. We apply this to a linear signal-response model providing analytical solutions, and to a nonlinear model of receptor-ligand systems. We show that the observational time $\tau$ has to be fine-tuned for an efficient detection of the irreversibility in time-series.
\end{abstract}

\maketitle


\section{Introduction}

The irreversibility of a process is the possibility to infer the existence of a time's arrow looking at an ensemble of realizations of its dynamics\cite{jarzynski2011equalities,parrondo2009entropy,feng2008length}.
This concept is formalized in thermodynamics as dissipated work or entropy production\cite{jarzynski1997nonequilibrium,crooks1999entropy,evans2002fluctuation}, a quantity that relates the probability of paths with their time-reversal conjugates\cite{kawai2007dissipation}.

Fluctuation theorems have been developed to describe the statistical properties of the entropy production and its relation to information-theoretic quantities in both Hamiltonian and Langevin dynamics\cite{jarzynski2000hamiltonian,chernyak2006path,ito2013information}.
Particular attention was given to measurement-feedback controlled models\cite{sagawa2012nonequilibrium,sagawa2010generalized} inspired by the Maxwell demon\cite{szilard1964decrease}, a gedanken-experiment in which physical work is extracted from thermodynamic systems using information. An information engine of this kind has been experimentally realized with a colloidal particle in an electric potential\cite{toyabe2010experimental}.

Here, we are interested in the stochastic dynamics of autonomous (uncontrolled) systems, where the irreversibility of trajectories results from nonconservative forces\cite{chernyak2006path}. 

We propose a generalization of the entropy production to the case of time-series resulting from a discretization with interval $\tau$ of continuous models. We call it \textit{mapping irreversibility} and we use the symbol $\varPhi_\tau$. Our motivation is a future use of the stochastic thermodynamics framework in the analysis of time-series data in biology and finance.

A key quantity here is the observational time $\tau$, and the detection of the irreversibility of processes is based on a fine-tuning of this parameter. We discuss this point with a model of receptor-ligand systems, where the entropy production measures the robustness of signaling.

We define signal-response models as continuous-time stationary processes characterized by the absence of feedback. In the bidimensional case they consist of a fluctuating signal $x$ and a dynamic response $y$. In a recent work\cite{auconi2017causal} we studied the information processing properties of linear multidimensional signal-response models and defined a measure of causal influence. Such definition is meaningful only in linear signal-response models.

The backward transfer entropy $T_{y\rightarrow x}(-\tau)$ is the standard transfer entropy\cite{cover2012elements} calculated in the ensemble of time-reversed trajectories, and it was shown to be related to the divergence of the dynamics from a hidden Markov model\cite{ito2016backward}.

On a bivariate framework with dynamic variables $x$ and $y$, we define the conditional entropy production $\varPhi_\tau^{y|x}$ of $y$ given $x$ with observational time $\tau$ (or conditional mapping irreversibility) as the difference between the entropy production of the two-dimensional system $(x,y)$ and the entropy production of the variable $x$ alone (with observational time $\tau$), $\varPhi_\tau^{y|x}=\varPhi_\tau^{xy}-\varPhi_\tau^x$.

We find an integral fluctuation theorem for signal-response models that involves the backward transfer entropy, and that is valid for any stationary (nonlinear) signal-response model. 
From this it follows the II law of thermodynamics for signal-response models, i.e. that the backward transfer entropy $T_{y\rightarrow x}(-\tau)$ of the response $y$ on the past of the signal $x$ is a lower bound to the conditional mapping irreversibility $\varPhi_\tau^{y|x}$:

\begin{eqnarray}\label{II_Law_NoFeedback}
\varPhi_\tau^{y|x}\geq T_{y\rightarrow x}(-\tau).
\end{eqnarray} 

This is our main result, and it shows the connection between the irreversibility of a process and the information flows towards the past between variables.
For the basic linear response model (BLRM discussed in \cite{auconi2017causal}), in the limit of small observational time $\tau$ the backward transfer entropy converges to the causal influence.

The paper is structured as follows. In section II.A we state the setting and formalism we use for the stochastic thermodynamics of time-series in the bivariate case, we motivate why we are interested in autonomous systems, we define the (conditional) mapping irreversibility, we introduce the spatial density of entropy production and we review the general integral fluctuation theorem\cite{seifert2012stochastic}. In section II.B we derive the fluctuation theorem for signal-response models involving the backward transfer entropy, and in section II.C we show how it applies to a dynamic linear signal-response model and to a biological model of receptor-ligand systems. In the Discussion section we review the results and we motivate further our definition of conditional mapping irreversibility comparing it with a possible alternative. We provide an Appendix section where the analytical solutions for the entropy production and for the backward transfer entropy in the BLRM are discussed.

\section{RESULTS}

\subsection{Bivariate time-series stochastic thermodynamics}

\subsubsection{Setting and definition of causal representations}
Let us consider an ensemble of trajectories generated by a continuous-time stochastic process composed of interacting variables stimulated by noise.
The stochastic differential equations describing such kind of systems can be represented as Bayesian networks, giving the probabilistic solution for the evolution as a function of the observational time interval $\tau$. Still, there are multiple ways of decomposing the joint probability distribution of states at the two instants $t$ and $t+\tau$. 
We say that a Bayesian network is a causal representation of the dynamics if conditional probabilities are expressed in a way that variables at time $t+\tau$ depend on variables at the same time instant or on variables at the previous time instant $t$ (and not vice-versa), and that the dependence structure is done in order to minimize the total number of conditions on the probabilities. This corresponds to minimizing the number of links in the Bayesian network describing the dynamics with observational time $\tau$. This is our preferred setting to describe information flows and causality\cite{auconi2017causal}, and to develop integral fluctuation theorems for the entropy production in stochastic thermodynamics.

In this paper we restrict ourselves to the bidimensional case with two stochastic dynamical variables $x$ and $y$. They have an interaction, described by the functions $g_x(x,y)$ and $g_y(x,y)$, which is in general asymmetric ($g_x(x,y)\neq \pm g_y(x,y)$). Taking Brownian motion $dW$ as noise stimulus, the stochastic differential equation for the general case is written in the Ito representation\citep{shreve2004stochastic} as:

\begin{eqnarray}\label{General}
\begin{cases} 
dx =g_x(x,y) dt + h_x(x,y)\sqrt{D_x}\,\, dW_x \\
dy =g_y(x,y) dt + h_y(x,y)\sqrt{D_y}\,\, dW_y
\end{cases}
\end{eqnarray}
where $D_x$ and $D_y$ are diffusion coefficients, and the $h$ functions are accounting for the case of multiplicative noise. Brownian motions are characterized by $\left\langle dW_i dW_j \right\rangle=\delta_{ij} dt$, for any $dt>0$.
 
We define the global variable $\zeta_\tau^{xy}$ as a couple of two successive states of system $(x,y)$ separated by a time interval $\tau$, $\zeta_\tau^{xy}\equiv(x(t)=x_t,y(t)=y_t,x(t+\tau)=x_{t+\tau},y(t+\tau)=y_{t+\tau})\equiv f_\tau^{xy}(x_t,y_t,x_{t+\tau},y_{t+\tau})\equiv(x_t,y_t,x_{t+\tau},y_{t+\tau})$. The functional form $f_\tau^{xy}$ is particularly convenient for the specification of the backward global variable $\widetilde{\zeta_{\tau}^{xy}}$. This is defined as the time-reversed conjugate of the global variable $\zeta_\tau^{xy}$, meaning the inverted couple of the same two successive states, $\widetilde{\zeta_{\tau}^{xy}}\equiv f_\tau^{xy}(x_{t+\tau},y_{t+\tau},x_t,y_t)\equiv(\widetilde{x_t},\widetilde{y_t},\widetilde{x_{t+\tau}},\widetilde{y_{t+\tau}})$. Correspondences of the type $\widetilde{x_t}=x_{t+\tau}$ are possible only when states at both times $t$ and $t+\tau$ are given. The probability density associated to the global variable $p(\zeta_{\tau}^{xy})$ is equal to the probability of the starting point $p(x_t,y_t)$ times the path-integral of the indicator of the trajectories that start and arrive in the states specified by $\zeta_{\tau}^{xy}$.

\subsubsection{Preliminary discussion of controlled systems}

A protocol\cite{jarzynski1997nonequilibrium} for influencing the $y$ dynamics controlling the $x$ variable, $\lambda(x)\equiv(x_t,x_{t+\tau})$, is defined here as a couple of successive states of the $x$ variable, $x_t,x_{t+\tau}$, that are kept fixed regardless of the dynamics of the $y$. Similarly to what is postulated by Ito-Sagawa for discrete-time dynamics on Bayesian networks\citep{ito2013information}, we assume that a detailed fluctuation theorem can be written for causal representations of time-series as a generalization of the detailed fluctuation theorem for single trajectories of controlled systems\citep{seifert2005entropy}. We define the entropy production with observational time $\tau$ (and we call it "mapping irreversibility") $\varphi^{y|\lambda(x)}_\tau $ of system $y$ controlled by system $x$ with protocol $\lambda(x)\equiv(x_t,x_{t+\tau})$ for a particular realization $(y_t,y_{t+\tau})$ as:

\begin{eqnarray}\label{detailed fluctuation theorem}
\varphi^{y|\lambda(x)}_\tau = \ln \left(\frac{p(y_t)}{p(y_{t+\tau})}\right) +\ln\left(\frac{p_\lambda(y_{t+\tau}|y_t,x_t,x_{t+\tau})}{p_\lambda(\widetilde{y_{t+\tau}}|\widetilde{y_t},\widetilde{x_t},\widetilde{x_{t+\tau}})}\right),~~~~~
\end{eqnarray} 
where the two terms on the right hand side correspond respectively to the entropy change of system $y$ and the entropy change in the thermal bath attached to $y$.

This is a generalization of the physical entropy production of continuous trajectories to the case of time-series. It describes the irreversibility of a transition in the stochastic map as a function of the observational time $\tau$. The physical entropy production rate of the original trajectory is found in the limit $\tau\rightarrow 0$.
With this definition for time-series (Eq.\ref{detailed fluctuation theorem}) we are able to treat more general systems of differential equations with mixed deterministic and stochastic variables, for which the continuous entropy production diverges. This is the case of Maxwell demon's deterministic strategies, when considering the demon as part of the system. In addition, we can now approach the data analysis of time-series in the framework of stochastic thermodynamics, where we needed a definition of entropy production that does not rely on a continuous-time limit. 

Note that the ensemble average $\braket{\varphi^{y|\lambda(x)}_\tau}$ of the entropy production is performed over the actual probability of the system $y$ and controlling variable $x$, $p(\zeta_\tau^{xy})$, where the control protocol $\lambda(x)\equiv(x_t,x_{t+\tau})$ may be influenced by the system $y$ dynamics. In $p_\lambda(y_{t+\tau}|y_t,x_t,x_{t+\tau})$ the control protocol $\lambda(x)$ fixes the two states of the controlling variable $x$ at times $t$ and $t+\tau$ as in $p(y_{t+\tau}|y_t,x_t,x_{t+\tau})$, but at intermediate instants the control variable $x$ can only follow trajectories that are compatible with the dynamics of system $x$ uninfluenced by $y$, that means in a different model where no measurement is performed and $y$ is not present in the differential equation for the evolution of $x$. This reduced dynamics without measurement is described by probability $p_\lambda(\zeta_\tau^{xy})\equiv p(\zeta_\tau^{y|\lambda(x)}) \cdot p_\lambda(x_t,x_{t+\tau})$.

A general fluctuation theorem connecting the entropy production in a thermodynamic system with the information used by controlling variables has been formulated in the case of fixed Bayesian networks\cite{ito2013information}. With the word "fixed" here we mean independent of a time discretization. There it is assumed that a fixed Bayesian network is the complete description of the dynamics, so that feedbacks between variables over time arise only from the structure of the directed graph.
This assumption was crucial because the detailed fluctuation theorem (Eq.\ref{detailed fluctuation theorem}) holds only in the absence of feedback control, and in general one would have to calculate the probabilities of transitions in both the real dynamics $p(y_{t+\tau}|y_t,x_t,x_{t+\tau})$ and the reduced dynamics $p_\lambda(y_{t+\tau}|y_t,x_t,x_{t+\tau})$, and for a fixed Bayesian network these two coincide.
In general, when the Bayesian network is a causal representation of a continuous-time dynamics, the two situations with and without feedback correspond to two different Bayesian networks, and the transition probabilities associated to these two different models are different, $p_\lambda(y_{t+\tau}|y_t,x_t,x_{t+\tau}) \neq p(y_{t+\tau}|y_t,x_t,x_{t+\tau})$, and also not related in a simple way. Therefore the general fluctuation theorem for fixed Bayesian networks\cite{ito2013information} does not hold for causal representations.

More explicitly, if we have knowledge of the state of a continuous time system $(x,y)$ at time $t$, that is $(x_t,y_t)$, then our estimate on the evolution of the variable $y$ at time $t+\tau$, $y_{t+\tau}$, would be influenced by a further knowledge on the evolution of the variable $x$ at time $t+\tau$, $x_{t+\tau}$, not only because $x$ is directly influencing $y$ ($x$ is present in the differential equation for $y$), but also because if $y$ is directly influencing $x$ ($y$ is present in the differential equation for $x$) then the knowledge of the $x$ transition $x_t,x_{t+\tau}$ gives information on $y$ at times in between $t$ and $t+\tau$ and therefore also on $y_{t+\tau}$. Then the probability $p(y_{t+\tau}|y_t,x_t,x_{t+\tau})$ is different from the same probability calculated in a model in which $y$ does not influence $x$ ($y$ is not present in the differential equation for $x$), $p_\lambda(y_{t+\tau}|y_t,x_t,x_{t+\tau})\neq p(y_{t+\tau}|y_t,x_t,x_{t+\tau})$.
We say that these two models, i.e. one with feedback and one without feedback, have two different causal representations. They correspond to two different Bayesian networks with different transition probabilities.

This is the case for most situations where Bayesian networks are built as a discretization of continuous time models, and this is also the case of measured time-series data from an experiment. For Langevin systems, it was shown\citep{ito2013information} that in the limit $\tau\rightarrow 0$ these two probabilities coverge to each other, and the fluctuation theorem for fixed Bayesian dynamics can be used for the entropy production rate.

We note that the Ito-Sagawa definition\cite{ito2013information} of entropy production for Bayesian networks differs from our Eq.\ref{detailed fluctuation theorem} also in the temporal order in which conditioning is preformed. This is discussed in the Discussion section and in Appendix C.

\subsubsection{The general fluctuation theorem on autonomous systems}

In this paper we study the probabilistic dynamics of systems in the absence of controlling variables. Our main interest is to relate the irreversibility of trajectories to the information flows between variables over time.

We define the mapping irreversibility, i.e. the stochastic entropy production with finite observational time $\tau$, for the autonomous uncontrolled system $(x,y)$ as a function of the global variable $\zeta_{\tau}^{xy}$ as:

\begin{eqnarray}\label{Definition}
\varphi_\tau^{xy}=\ln \left(\frac{p(\zeta_\tau^{xy})}{p(\widetilde{\zeta_{\tau}^{xy}})}\right).
\end{eqnarray}

This quantity satisfies the integral fluctuation theorem\cite{seifert2012stochastic}, i.e.:

\begin{eqnarray}\label{IFT}
\left\langle e^{-\varphi_\tau^{xy}}\right\rangle_{p(\zeta_\tau^{xy})}=\int_\Omega d\zeta_{\tau}^{xy}  p(\widetilde{\zeta_{\tau}^{xy}})=1,
\end{eqnarray}
where $d\zeta_{\tau}^{xy}=d x_t d y_t d x_{t+\tau} d y_{t+\tau}$, $\widetilde{d x_t}=d x_{t+\tau}$, and $\Omega$ is the whole space of the global variable.

From the convexity of the exponential function it follows that the entropy production $\varPhi_\tau^{xy}$, which is the ensemble average of the stochastic entropy production $\varphi_\tau^{xy}$, is non-negative. This is the II law of Thermodynamics for the bivariate system $(x,y)$:
\begin{eqnarray}\label{II_Law}
\varPhi_\tau^{xy}=\left\langle \varphi_\tau^{xy}\right\rangle _{p(\zeta_\tau^{xy})}\geq 0.
\end{eqnarray}

We define the conditional mapping irreversibility of $y$ given $x$ as the difference between the mapping irreversibility of the system $(x,y)$ and the mapping irreversibility of system $x$ alone:

\begin{eqnarray}\label{Conditional_entropy_production}
\varPhi_\tau^{y|x}=\varPhi_\tau^{xy}-\varPhi_\tau^{x}=\left\langle\ln\left(\frac{p(y_t,y_{t+\tau}|x_t,x_{t+\tau})}{p(\widetilde{y_t},\widetilde{y_{t+\tau}}|\widetilde{x_t},\widetilde{x_{t+\tau}})}\right)\right\rangle_{p(\zeta_\tau^{xy})}.~~~~
\end{eqnarray}

\begin{figure}
	\begin{center}
		\includegraphics[scale=0.32]{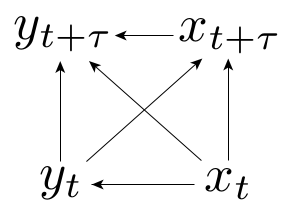}
		\caption{\label{graph_complete} Complete causal representation. The arrows represent the way we decompose the joint probability density. In the complete case we have $p(\zeta_\tau^{xy})=p(x_t)\cdot p(y_t|x_t)\cdot p(x_{t+\tau}|x_t,y_t)\cdot p(y_{t+\tau}|x_t,y_t,x_{t+\tau})$.}
	\end{center}
\end{figure}

In the general case of a complete causal representation resulting from the dynamics (Fig.\ref{graph_complete}), where all edges are present in the Bayesian network, nothing more can be said than the II of Thermodynamics (Eq.\ref{II_Law}). As an example, this happens when the evolution of each variable is influenced by all other variables (Eq.\ref{General}).

We argue that more informative fluctuation theorems arise as a consequence of missing edges in the causal representation of the dynamics in terms of Bayesian networks. In the bivariate case there is only one class of continuous-time models for which informative fluctuation theorems for causal representations can be written: the signal-response models.

\subsubsection{The spatial density of entropy production}

Let us also use an equivalent representation of the entropy production in terms of backward probabilities\cite{ito2016information} defined as $p_B(\zeta_\tau^{xy})=p(x(t)=x_t,y(t)=y_t,x(t-\tau)=x_{t+\tau},y(t-\tau)=y_{t+\tau})$. For stationary processes it holds $p_B(\zeta_\tau^{xy})=p(\widetilde{\zeta_{\tau}^{xy}})$ and $\varphi^{xy}_\tau=\ln(\frac{p(\zeta_\tau^{xy})}{p_B(\zeta_\tau^{xy})})$.

We introduce here the spatial density of entropy production (with observatinal time $\tau$) for stationary processes as:

\begin{eqnarray}\label{spatial} 
&\psi(x_t,y_t)=\int^\infty _{-\infty}\int^\infty _{-\infty} dx_{t+\tau} dy_{t+\tau} p(\zeta_\tau^{xy}) \varphi^{xy}_\tau =\nonumber \\&=\int^\infty _{-\infty}\int^\infty _{-\infty} dx_{t+\tau} dy_{t+\tau} p(\zeta_\tau^{xy})\ln\left(\frac{p(\zeta_\tau^{xy})}{p_B(\zeta_\tau^{xy})}\right)\nonumber \\&=  p(x_t,y_t) \int^\infty _{-\infty}\int^\infty _{-\infty} dx_{t+\tau} dy_{t+\tau}  p(x_{t+\tau},y_{t+\tau}|x_t,y_t) * \nonumber \\& *\ln\left(\frac{p(x(t+\tau)=x_{t+\tau},y(t+\tau)=y_{t+\tau}|x(t)=x_t,y(t)=y_t)}{p(x(t-\tau)=x_{t+\tau},y(t-\tau)=y_{t+\tau}|x(t)=x_t,y(t)=y_t)}\right) 
\end{eqnarray}

The spatial density of entropy production $\psi(x_t,y_t)$ tells us which situations $(x_t,y_t)$ contribute more to the irreversibility of the macroscopic process. $\psi(x_t,y_t)$ is proportional to the distance (precisely to the Kullback–Leibler divergence\cite{cover2012elements}) of the distribution of future states $p(x_{t+\tau},y_{t+\tau}|x_t,y_t)$ to the distribution of past states $p(x_{t-\tau},y_{t-\tau}|x_t,y_t)$ of the same condition $(x_t,y_t)$.

\subsection{The fluctuation theorem for signal-response models}

\begin{figure}
	\begin{center}
		\includegraphics[scale=0.32]{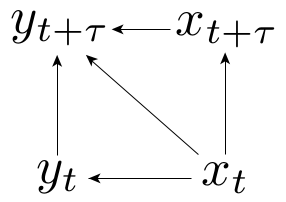}
		\caption{\label{graph_SRM} Causal representation of signal-response models. The joint probability density is decomposed into $p(\zeta_\tau^{xy})=p(x_t)\cdot p(y_t|x_t)\cdot p(x_{t+\tau}|x_t)\cdot p(y_{t+\tau}|x_t,y_t,x_{t+\tau})$.}
	\end{center}
\end{figure}

If the system $(x,y)$ is such that the variable $y$ does not influence the dynamics of the variable $x$, then we are dealing with signal-response models (Fig.\ref{graph_SRM}). The stochastic differential equation for signal-response models is written in the Ito representation\citep{shreve2004stochastic} as:

\begin{eqnarray}\label{GeneralNoFeedback}
\begin{cases}
dx =g_x(x) dt + h_x(x)\sqrt{D_x}\,\, dW_x \\
dy =g_y(x,y) dt + h_y(x,y)\sqrt{D_y}\,\, dW_y
\end{cases}
\end{eqnarray}
The absence of feedback is written in $\frac{\partial g_x}{\partial y}=\frac{\partial h_x}{\partial y}=0$. As a consequence the conditional probability satisfies $p(y_t|x_t,x_{t+\tau})=p(y_t|x_t)$, and the corresponding causal representation is incomplete, see the Bayesian network in Fig.\ref{graph_SRM}.

For signal-response models we can provide a lower bound on the entropy production that is more informative than Eq.\ref{II_Law}, and that involves the backward transfer entropy $T_{y\rightarrow x}(-\tau)$. The backward transfer entropy\cite{ito2016backward} is defined as the standard transfer entropy for the ensemble of time-reversed trajectories. The stochastic counterpart as a function of $\zeta_{\tau}^{xy}\setminus y_t$ is defined as:

\begin{eqnarray}\label{BTE}
T^{st}_{y\rightarrow x}(-\tau)=\ln\left(\frac{p(x_t|y_{t+\tau},x_{t+\tau})}{p(x_t|x_{t+\tau})}\right),
\end{eqnarray}
where $st$ stands for stochastic.

Then by definition $T_{y\rightarrow x}(-\tau)=\left\langle T^{st}_{y\rightarrow x}(-\tau)\right\rangle_{p(\zeta_\tau^{xy})}$.
We keep the same symbol $T_{y\rightarrow x}$ as the standard transfer entropy because in stationary processes the backward transfer entropy is the standard transfer entropy (calculated on forward trajectories) for negative shifts $-\tau$. It measures information flows towards past, and in stationary processes it does not depend on the instant $t$ but only on the observational time $\tau$.

The fluctuation theorem for signal-response models is:

\begin{eqnarray}\label{TheoremNoFeedback}
&\left\langle e^{-\varphi_\tau^{xy}+\varphi_\tau^{x}+T^{st}_{y\rightarrow x}(-\tau)}\right\rangle _{p(\zeta_\tau^{xy})}=\nonumber\\&=\int_\Omega d\zeta_{\tau}^{xy} p(\widetilde{y_{t+\tau}}|\widetilde{x_t},\widetilde{y_t},\widetilde{x_{t+\tau}}) p(x_t,x_{t+\tau},y_{t+\tau}) =1,~~~~
\end{eqnarray}
where we used the signal-response property of no feedback $p(\widetilde{y_t}|\widetilde{x_t},\widetilde{x_{t+\tau}})=p(\widetilde{y_t}|\widetilde{x_t})$, the correspondence $dy_t=\widetilde{dy_{t+\tau}}$, and the normalization $\int^\infty_{-\infty} d \widetilde{y_{t+\tau}} p(\widetilde{y_{t+\tau}}|\widetilde{x_t},\widetilde{y_t},\widetilde{x_{t+\tau}})=1$.

From the convexity of the exponential it follows the II law for signal-response models (Eq.\ref{II_Law_NoFeedback}):

\begin{eqnarray}\nonumber
\varPhi_\tau^{y|x}=\varPhi_\tau^{xy}-\varPhi_\tau^{x}\geq T_{y\rightarrow x}(-\tau),
\end{eqnarray}

which is the main result of our paper.

\subsection{Applications}

\subsubsection{The basic linear response model}

We study the II law for signal-response models (Eq.\ref{II_Law_NoFeedback}) in the basic linear response model (BLRM), whose information processing properties for the forward trajectories are already discussed in \citep{auconi2017causal}. The BLRM is composed of a fluctuating signal $x$ described by the Ornstein-Uhlenbeck process\citep{uhlenbeck1930theory,gillespie1996exact}, and a dynamic linear response $y$ to this signal:

\begin{eqnarray}\label{BLRM}
\begin{cases} 
dx =-\frac{x}{t_{rel}} dt + \sqrt{D}\,\, dW \\
\frac{dy}{dt}=\alpha x -\beta y
\end{cases}
\end{eqnarray}

Note that the response $y$ is not coupled to a thermal bath, $D_y=0$, while the signal is, $D_x= D>0$. For any finite interval $\tau$, this corresponds to the limit of weak coupling $D_y\rightarrow 0$.

This model allows analytical representations for the mapping irreversibility $\varPhi_\tau^{xy}$ (calculated in Appendix A) and the backward transfer entropy $T_{y\rightarrow x}(-\tau)$ (calculated in Appendix B). We find that, once the observational time $\tau$ is specified, $\varPhi_\tau^{xy}$ and $T_{y\rightarrow x}(-\tau)$ are both functions of just the two parameters $t_{rel}$ and $\beta$, which describe respectively the time scale of the fluctuations of the signal and the time scale of the response to a deterministic input.

Since the signal is a time-symmetric (reversible) process, $\varPhi_\tau ^x=0$, the backward transfer entropy $T_{y\rightarrow x}(-\tau)$ is the lower bound on the total entropy production $\varPhi_\tau^{xy}$ in the BLRM.

\begin{figure}
	\begin{center}
		\includegraphics[scale=0.235]{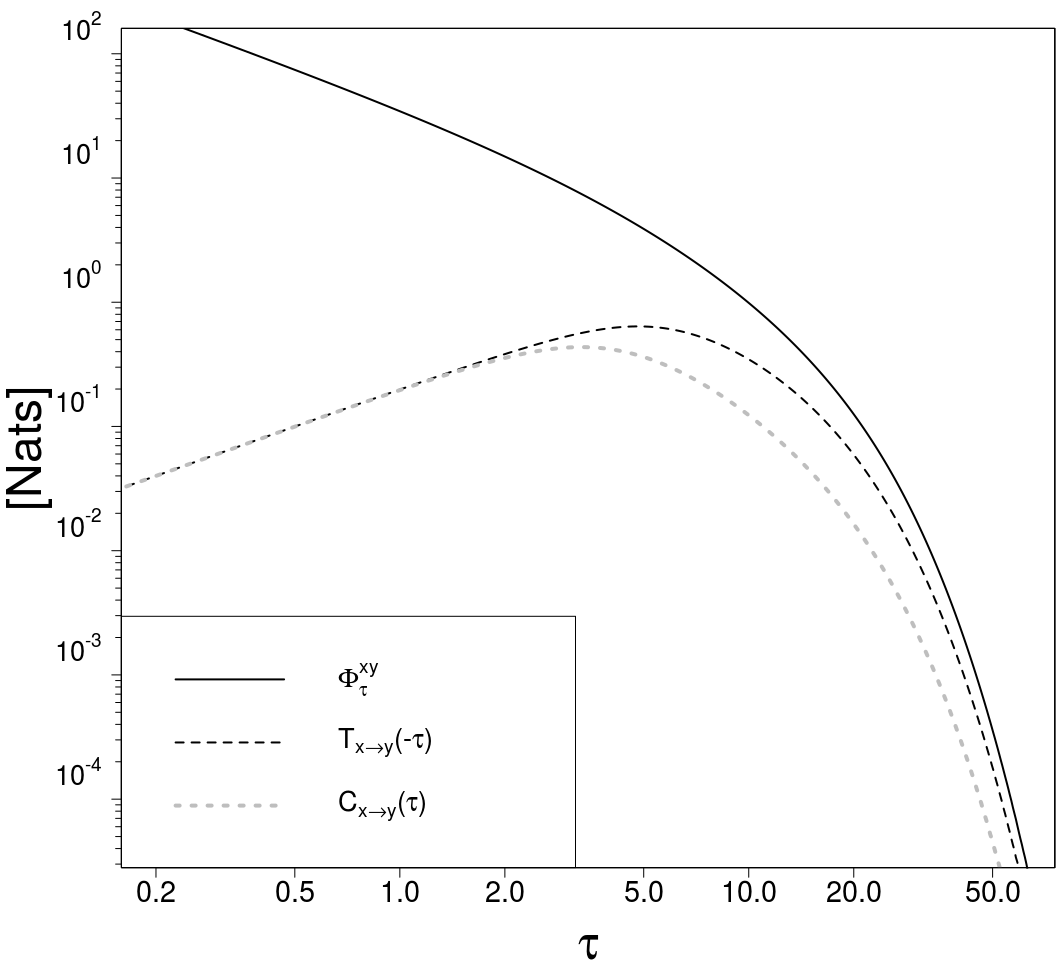}
		\caption{\label{One} Mapping irreversibility $\varPhi_\tau^{xy}$, backward transfer entropy $T_{y\rightarrow x}(-\tau)$ and causal influence $C_{x\rightarrow y}(\tau)$ in the the BLRM as a function of the observational time interval $\tau$. The parameters are $\beta=0.2$ and $t_{rel}=10$. All graphs are produced using R\cite{team2014r}.}
	\end{center}
\end{figure}

The plot in Fig.\ref{One} shows the mapping irreversibility $\varPhi_\tau^{xy}$ and the backward transfer entropy $T_{y\rightarrow x}(-\tau)$ as a function of the observational time $\tau$. In the limit of small $\tau$, the entropy production diverges because of the deterministic nature of the response dynamics (the standard deviation on the determination of the velocity $\frac{dy}{dt}$ due to instantaneous movements of the signal vanishes as $\alpha \sqrt{D} \sqrt{dt}\rightarrow 0$). The backward transfer entropy $T_{y\rightarrow x}(-\tau)$ instead vanishes for $\tau\rightarrow 0$ because the Brownian motion has nonzero quadratic variation\cite{shreve2004stochastic} and is the dominating term in the signal dynamics for small time intervals.
In the limit of large observational time intervals $\tau\rightarrow \infty$ the entropy production is asymptotically double the backward transfer entropy, that is its lower bound given by the II law for signal-response models (Eq.\ref{II_Law_NoFeedback}), $\frac{\varPhi_\tau^{xy}}{T_{y\rightarrow x}(-\tau)}\rightarrow 2$. This limit of $2$ is valid for any choice of the parameters in the BLRM.

Interestingly, for small observational time $\tau\rightarrow 0$, the causal influence of the signal on the evolution of response (defined in \citep{auconi2017causal}) converges to the backward transfer entropy of the response on the past of the signal $C_{x\rightarrow y}(\tau)\rightarrow T_{y\rightarrow x}(-\tau)$. For large observational time $\tau\rightarrow \infty$ instead the causal influence converges to the standard (forward) transfer entropy $C_{x\rightarrow y}(\tau)\rightarrow T_{y\rightarrow x}(\tau)$. Also in this limit $\tau\rightarrow \infty$, the causal influence is an eighth of the entropy production $\frac{\varPhi_\tau^{xy}}{C_{x\rightarrow y}(\tau)}\rightarrow 8$ for any choice of the parameters in the BLRM. 

\begin{figure}
	\begin{center}
		\includegraphics[scale=0.22]{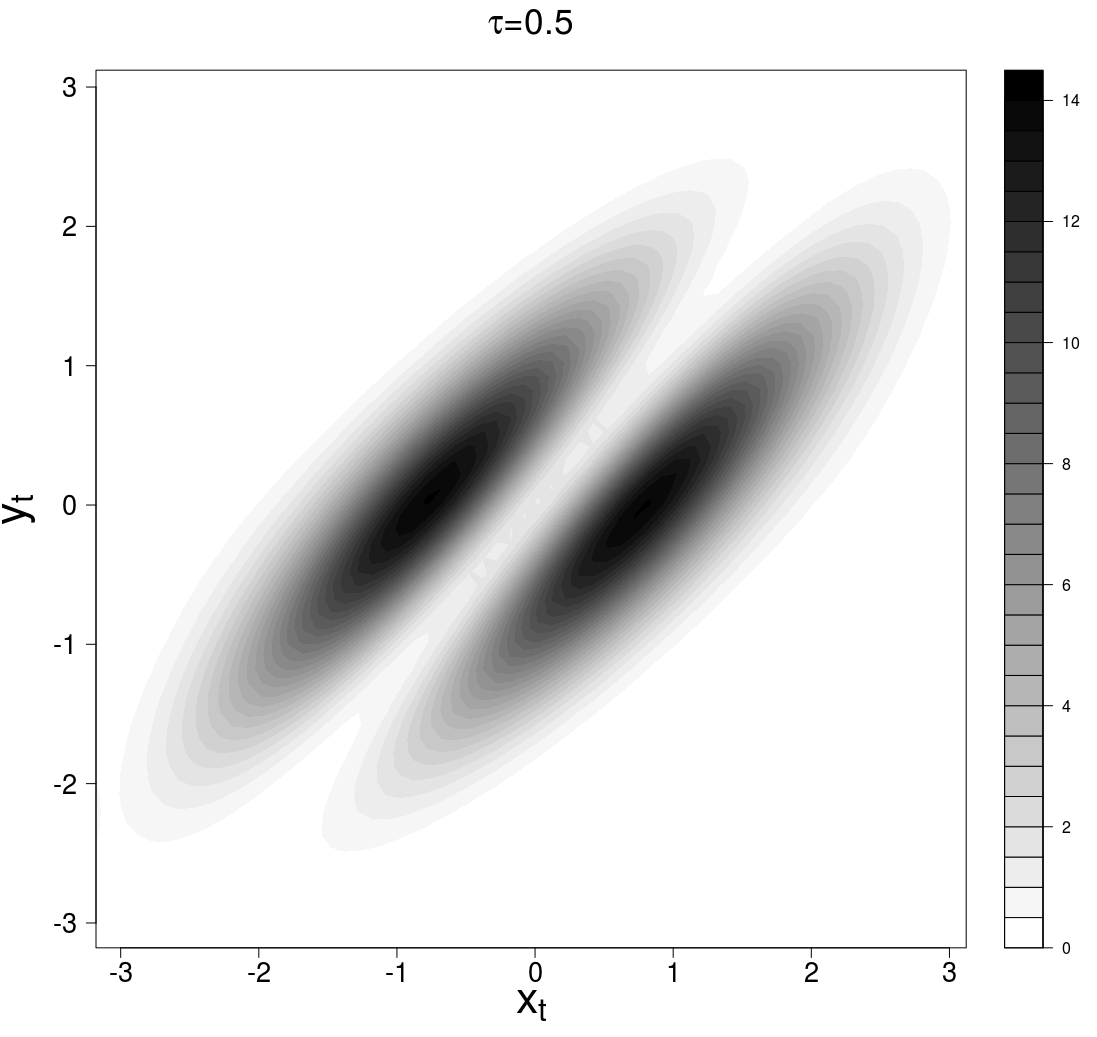}
		\caption{\label{Two} Spatial density of entropy production $\psi(x_t,y_t)$ for the BLRM at $\tau=0.5<\frac{1}{\beta}<t_{rel}$. The parameters are $\beta=0.2$ and $t_{rel}=10$. Both $\psi(x_t,y_t)$ and the space $(x,y)$ are expressed in units of the standard deviation of the dynamics.}
	\end{center}
\end{figure}

\begin{figure}
	\begin{center}
		\includegraphics[scale=0.22]{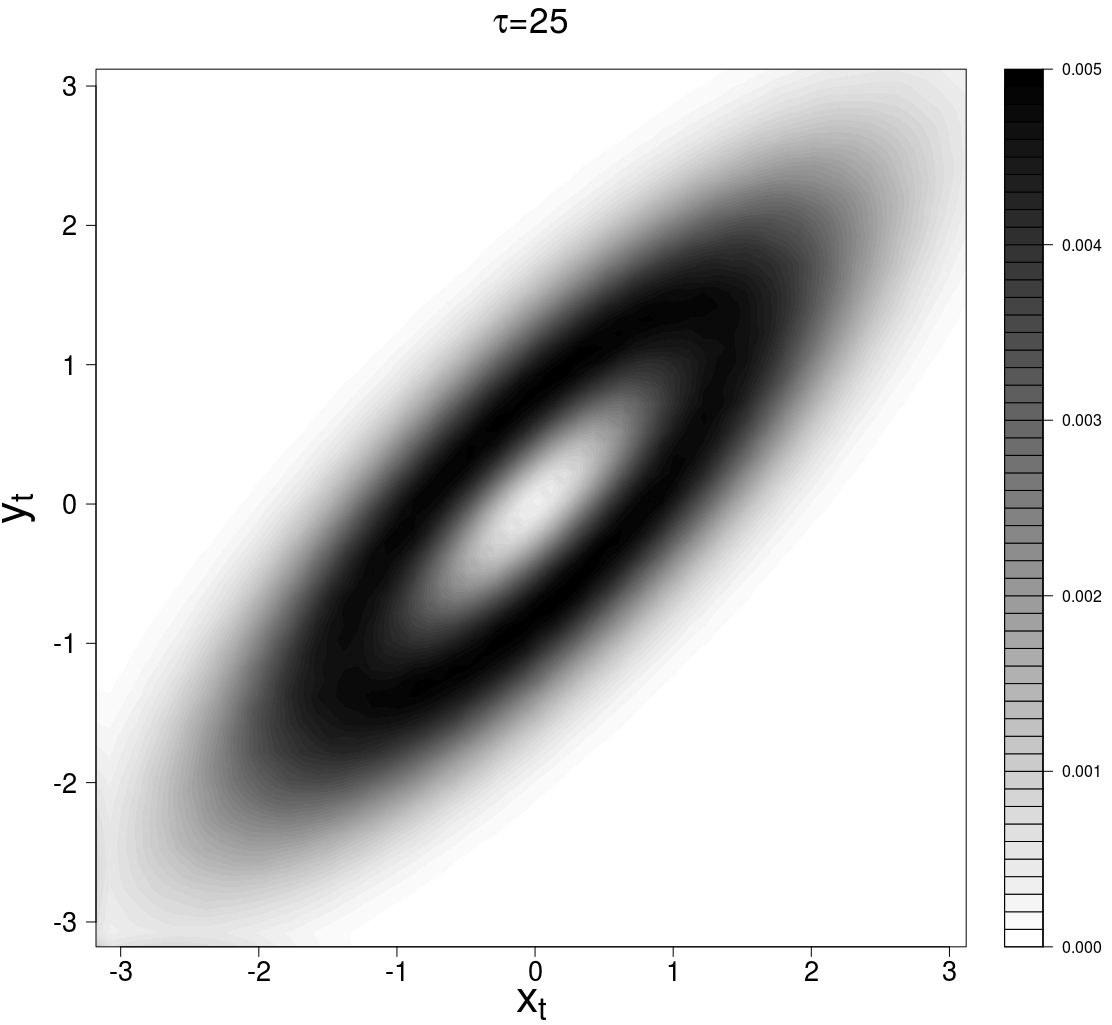}
		\caption{\label{Three} Spatial density of entropy production $\psi(x_t,y_t)$ for the BLRM at $\tau=25>t_{rel}>\frac{1}{\beta}$. The parameters are $\beta=0.2$ and $t_{rel}=10$. Both $\psi(x_t,y_t)$ and the space $(x,y)$ are expressed in units of the standard deviation of the dynamics.}
	\end{center}
\end{figure}

Let us now consider the spatial density of entropy production $\psi(x_t,y_t)$ in the BLRM for small and large observational time $\tau$. In Fig.\ref{Two} we choose a $\tau$ smaller than the characteristic response time $\frac{1}{\beta}$ and also smaller than the characteristic time of fluctuations $t_{rel}$. In the limit $\tau\rightarrow 0$ the signal dynamics is dominated by noise and the entropy production is mainly given by movements of the response $y$. The two spots correspond to the points where the product of the density $p(x_t,y_t)$ times the absolute value of the instant velocity $\dot{y}$ is larger. For longer intervals $\tau\gtrapprox\frac{1}{\beta}$ (that is the case of Fig.\ref{Three}) the history of the signal becomes relevant because it determined the present value of the response, therefore the irreversibility density is also distributed on those points of the diagonal (corresponding to roughly $\dot{y}=0$) where the absolute value of the response $y$ is big enough. Also as a consequence, in this regime the backward transfer entropy is a meaningful lower bound on the entropy production, that is Eq.\ref{II_Law_NoFeedback}.

\subsubsection{Receptor-ligand systems}

The Receptor-Ligand interaction is the fundamental mechanism of molecular recognition in biology and is a recurring motif in signaling pathways\cite{klipp2016systems,kholodenko2006cell}. The fraction of activated receptors is part of the cell's representation of the outside world, it is the cell's estimate on the concentration of ligands in the environment, upon which it bases its protein expression and response to external stress.

If we could experimentally keep the concentration of ligands fixed we would still get a fluctuating number of activated receptors due to the intrinsic stochasticity of the macroscopic description of chemical reactions. 
Recent studies allowed a theoretical understanding of the origins of the macroscopic "noise" (i.e. the output variance in the conditional probability distributions), and also raised questions about the optimality of the input distributions in terms of information transmission\cite{bialek2005physical,tkavcik2008information,crisanti2018statistics,waltermann2011information}.

Here we consider the dynamical aspects of information processing in receptor-ligand systems\cite{di2012short,nemenman2012gain}, where the response is integrated over time. 
If the perturbation of the receptor-ligand binding on the concentration of free ligands is negligible, that is in the limit of high ligand concentration, receptor-ligand systems can be modeled as nonlinear signal-response models\cite{di2014simple}. We write our model of receptor-ligand systems in the Ito representation\cite{shreve2004stochastic} as: 

\begin{eqnarray}\label{RLM}
\begin{cases} 
dx =-(x-1) dt + x\, dW_x \\
dy =k_{on}(1-y) \frac{x^h}{1+x^h} dt -k_{off} y dt+ y(1-y) dW_y~~~
\end{cases}
\end{eqnarray} 

The fluctuations of the ligand concentration $x$ are described by a mean-reverting geometric Brownian motion, with an average $\braket{x}=1$ in arbitrary units. The response, that is the fraction of activated receptors $y$, is driven by a Hill-type interaction with the signal with cooperativity coefficient $h$, and chemical bound/unbound rates $k_{on}$ and $k_{off}$. For simplicity, the dynamic range of the response is set to be coincident with the mean value of the ligand concentration, that means to choose a Hill constant $K=\braket{x}=1$. The form of the $y$ noise is set by the biological constraint $0<y<1$. For simplicity, we choose a cooperativity coefficient of $h=2$, that is the lower order of sigmoidal functions.

The mutual information between the concentration of ligands and the fraction of activated receptors in a cell is a natural choice for quantifying its sensory properties\cite{tkavcik2009optimizing}. Here we argue that, in the case of signal-response models, the conditional entropy production is the relevant measure, because it quantifies how the dynamics of the signal produces irreversible transitions in the dynamics of the response, which is closely related to the concept of causation. Besides, our measure of causal influence\cite{auconi2017causal} has yet not been generalized to the nonlinear case, while the entropy production has a consistent thermodynamical interpretation\cite{seifert2012stochastic}.

\begin{figure}
	\begin{center}
		\includegraphics[scale=0.235]{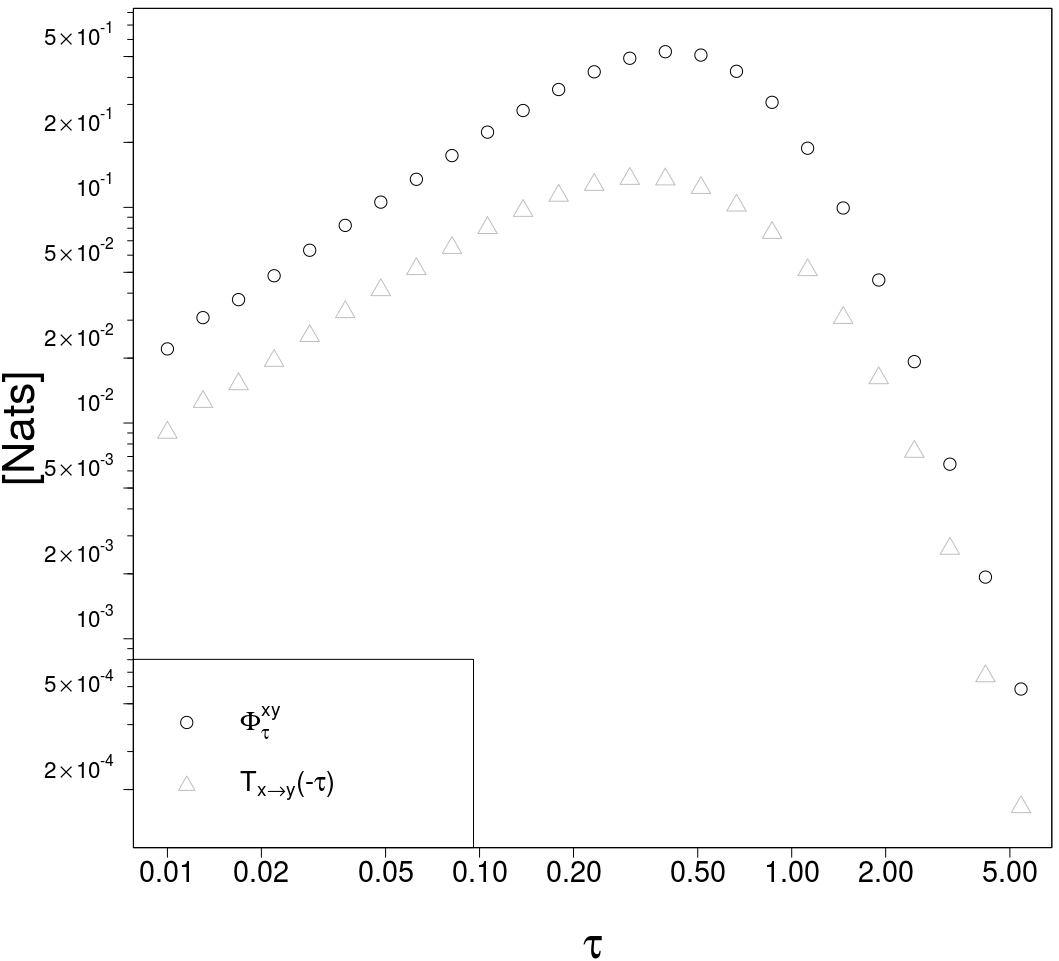}
		\caption{\label{Fig_RLM} Entropy production and backward transfer entropy in our model of receptor-ligand systems (Eq.\ref{RLM}). The parameters are $k_{on}=5$, $k_{off}=1$, $h=2$ and $t_{rel}=10$.}
	\end{center}
\end{figure}

We simulated the receptor-ligand model of Eq.\ref{RLM}, and we evaluated numerically the mapping irreversibility $\varPhi_\tau^{xy}$ and the backward transfer entropy $T_{y\rightarrow x}(-\tau)$ using a multivariate Gaussian approximation for the conditional probabilities $p(x_{t+\tau},y_{t+\tau}|x_t,y_t)$ (details in Appendix D). The II law for signal response models sets $\varPhi_\tau^{xy}\geq T_{y\rightarrow x}(-\tau)$ and proves to be a useful tool for receptor-ligand systems, as it is seen if Fig.\ref{Fig_RLM}. Note that the numerical estimation of the entropy production requires many more samples compared to the backward transfer entropy: $\varPhi_\tau^{xy}$ depends on $\zeta_{\tau}^{xy}$ (4 dimensions) while $T_{y\rightarrow x}(-\tau)$ depends on $\zeta_{\tau}^{xy}\setminus y_t$ (3 dimensions). In a real biological experimental setting the sampling process is expensive, and the backward transfer entropy is therefore a useful lower bound for the entropy production, and an interesting characterization of the system to be used when the number of samples is not large enough.

The intrinsic noise of the response $y(1-y)dW_y$ is the dominant term in the response dynamics for small intervals $\tau$. This makes both $\varPhi_\tau^{xy}$ and $T_{y\rightarrow x}(-\tau)$ vanish in the limit $\tau\rightarrow 0$. In the limit of large observational time $\tau$, as it is also the case for the BLRM and in any stationary process, the entropy production for the corresponding time-series $\varPhi_\tau^{xy}$ and all the information measures are vanishing, because the memory of the system is damped exponentially over time by the relaxation parameter $k_{off}$ ($\beta$ in the BLRM). Therefore in order to better detect the irreversibility of a process one must choose an appropriate observational time $\tau$. In the receptor-ligand model of Eq.\ref{RLM} with parameters $k_{on}=5$, $k_{off}=1$, $h=2$ and $t_{rel}=10$ we see that the optimal observational time is around $\tau\approx0.5$ (see Fig.\ref{Fig_RLM}). Here for "optimal" we mean the observational time that corresponds to the highest mapping irreversibility $\varPhi_\tau^{xy}$, but one might also be interested in inferring the entropy rate (that is $\frac{\varPhi_\tau^{xy}}{\tau}$ in the limit $\tau\rightarrow 0$) looking at time-series data with finite sampling interval $\tau$. We do not treat this problem here.

\section{DISCUSSION}

This work is based on our definition of mapping irreversibility (Eq.\ref{Definition}), that is the entropy production of time-series obtained from a discretization with observational time $\tau$ of continuous dynamics.

For controlled models the entropy production takes the form of Eq.\ref{detailed fluctuation theorem}. This definition is different from the one used by Ito-Sagawa\citep{ito2013information,ito2016information}. Their alternative definition of discrete entropy production of system $y$ controlled by protocol $\lambda(x)$ in a single time step is:

\begin{eqnarray}\label{Ito-Sagawa} 
\eta^{y|\lambda(x)}_\tau =\ln\left(\frac{p(y_t)}{p(y_{t+\tau})}\right)+ \ln\left(\frac{p_\lambda(y_{t+\tau}|y_t,x_t,x_{t+\tau})}{p_\lambda(\widetilde{y_{t+\tau}}|\widetilde{y_t},x_t,x_{t+\tau})}\right).~~~~~~
\end{eqnarray}

When feedbacks enter the dynamics at shorter scales compared to the observational time $\tau$, as it is the case in our setting of causal representations, the transition probabilities in the reduced dynamics (where no feedback is performed) $p_\lambda(y_{t+\tau}|y_t,x_t,x_{t+\tau})$ are different compared to the transition probabilities in the original dynamics $p(y_{t+\tau}|y_t,x_t,x_{t+\tau})$. This is not problematic in their setting though, because in a fixed Bayesian network the feedbacks arise as edges paths in the network and are not deriving from an underlying continuous dynamics.

The difference with our definition (Eq.\ref{detailed fluctuation theorem}) is in the way the discrete protocol $\lambda(x)$ is applied. They perform the conditioning for the backward transitions on the same exact states as in the forward transitions. In our setting of bivariate causal representations, their conditioning for the backward transitions would be on the original states $(x_t,x_{t+\tau})$ rather than on the time reversal conjugates $(\widetilde{x_t},\widetilde{x_{t+\tau}})$ (compare Eq.\ref{detailed fluctuation theorem} and Eq.\ref{Ito-Sagawa}). This difference becomes evident in the limit $\tau\rightarrow \infty$ as we discuss in Appendix C for the BLRM. There, where no feedback on the signal is performed by the response, the alternative irreversibility (Eq.\ref{Ito-Sagawa}) has a finite limit (instead of zero) for large time intervals $\tau\rightarrow \infty$,  $\left\langle\eta^{y|\lambda(x)}_\tau\right\rangle _{p(\zeta_\tau^{xy})} \rightarrow \beta t_{rel}$. Note that in this limit the forward time-series are indistinguishable from their time-reversed conjugates, and $\left\langle\eta^{y|\lambda(x)}_\tau\right\rangle _{p(\zeta_\tau^{xy})}$ is greater than zero because of the correlation $\braket{xy}$ being greater than zero. In addition, we note that because of this way of conditioning, the term $\eta^{y|\lambda(x)}_\tau-\ln\left(\frac{p(y_t)}{p(y_{t+\tau})}\right)$ does not have a simple relation to the total entropy production of the system $(x,y)$ like our Eq.\ref{Conditional_entropy_production}.

A modification of the alternative definition $\eta^{y|\lambda(x)}_\tau$ in which the reduced dynamics dependence is removed, can be written as:

\begin{eqnarray}\label{modified Ito-Sagawa} 
\eta^{y|x}_\tau = \ln\left(\frac{p(y_{t+\tau}|y_t,x_t,x_{t+\tau})}{p(\widetilde{y_{t+\tau}}|\widetilde{y_t},x_t,x_{t+\tau})}\right)+\ln\left(\frac{p(y_t)}{p(y_{t+\tau})}\right).~~~~~
\end{eqnarray}

For this modified version, a fluctuation theorem can be written, and it is the Ito-Sagawa fluctuation theorem proved in \citep{ito2013information}. In stationary systems it involves the difference between forward and backward transfer entropies:

\begin{eqnarray}\label{TheoremItoSagawa}
&\braket{e^{-\eta_\tau^{y|x}+T^{st}_{y\rightarrow x}(-\tau)-T^{st}_{y\rightarrow x}(\tau)}}_{p(\zeta_\tau^{xy})}=1.
\end{eqnarray}

The two different definitions for controlled systems Eq.\ref{detailed fluctuation theorem} and Eq.\ref{Ito-Sagawa}, with different ways of applying the conditioning on the protocol, still converge for Langevin systems in the limit $\tau\rightarrow 0$, because the discretization scheme enters the entropy production with terms that are vanishing faster than $\tau$ as it is shown in \citep{ito2013information}. 

Nevertheless, we argue that in the more general case of time-series the Ito-Sagawa definition should be modified into Eq.\ref{detailed fluctuation theorem} for controlled system and to Eq.\ref{Definition} for autonomous systems.

We used our definition to study the irreversibility of stochastic maps resulting from a time discretization with observational time $\tau$ of continuous models.
While for autonomous systems in the general case the only statement we can provide is the II law of thermodynamics (Eq.\ref{II_Law}), a more informative lower bound on the entropy production is found for signal-response models (Eq.\ref{II_Law_NoFeedback}). This sets the backward transfer entropy as a lower bound to the entropy production, and describes the connection between the irreversibility of stochastic trajectories and the information flows towards past between variables.

We restrict ourselves to the bivariate case here, but we conjecture that fluctuation theorems for multidimensional stochastic autonomous dynamics should arise in general as a consequence of missing arrows in the (non complete, see e.g. Fig.\ref{graph_SRM}) causal representation of the dynamics in terms of Bayesian networks.

In our opinion, a general relation connecting the incompleteness of the causal representation of the dynamics and fluctuation theorems is still lacking.

We also introduced a discussion about the observational time $\tau$ in data analysis. In a biological model of receptor-ligand systems we showed that it has to be fine-tuned for a robust detection of the irreversibility of the process, which is related to the concept of causation\citep{auconi2017causal} and therefore to the efficiency of the biological coupling between signalling and response.\\

\section*{Acknowledgements}
We thank Wolfgang Giese, Valentina T Tovazzi, Friedemann Uschner, Bj\"orn Goldenbogen, Ana Bulovic, Roman Rainer, Matthias Reis and Martin Seeger for useful discussions. We thank Marco Scazzocchio for numerical methods and algorithms expertise. We thank Jesper Romers for discussions of mathematical aspects.

Work at Humboldt-Universit\"at zu Berlin was supported by the DFG (Graduiertenkolleg 1772 for Computational Systems Biology).

\begin{widetext}
\section*{APPENDIX A: MAPPING IRREVERSIBILITY IN THE BLRM}

Let us consider an ensemble of stochastic trajectories generated with the BLRM (Eq.\ref{BLRM}). The mapping irreversibility $\varPhi_\tau^{xy}$ here is the Kullback-Leibler divergence\citep{cover2012elements} between the probability density $p(\zeta_{\tau}^{xy})$ of couples of successive states $\zeta_{\tau}^{xy}$ separated by a time interval $\tau$ of the original trajectory and the probability density $p_B(\zeta_{\tau}^{xy})=p(\widetilde{\zeta_{\tau}^{xy}})$ of the same couples of successive states $\zeta_{\tau}^{xy}$ of the time-reversed conjugate of the original trajectory (Eq.\ref{Definition}). For the sake of clarity, we use here in this appendix the full formalism rather than the compact one based on the functional form $f_\tau^{xy}$. 

The time-reversed density of a particular couple of successive states, $(x(t)=\gamma,y(t)=\delta)$ and $(x(t+\tau)=\mu,y(t+\tau)=\xi)$, is equivalent to the original density of the exchanged couple of states, $(x(t)=\mu,y(t)=\xi)$ and $(x(t+\tau)=\gamma,y(t+\tau)=\delta)$. Therefore the density $p(\widetilde{\zeta_{\tau}^{xy}})=p(x(t)=\mu,y(t)=\xi,x(t+\tau)=\gamma,y(t+\tau)=\delta)$ is the transpose of the density $p(\zeta_{\tau}^{xy})=p(x(t)=\gamma,y(t)=\delta,x(t+\tau)=\mu,y(t+\tau)=\xi)$.

The mapping irreversibility for the BLRM is then written as:

\begin{eqnarray}\label{A1}
&\varPhi_\tau^{xy}=\braket{\varphi_\tau^{xy}}_{p(\zeta_\tau^{xy})}= \int_{-\infty}^{\infty}\int_{-\infty}^{\infty}\int_{-\infty}^{\infty}\int_{-\infty}^{\infty} d\gamma d\delta d\mu d\xi~ p\left(x(t)=\gamma,y(t)=\delta,x(t+\tau)=\mu,y(t+\tau)=\xi\right) *\nonumber\\ &*\ln\left(\frac{p(x(t)=\gamma,y(t)=\delta,x(t+\tau)=\mu,y(t+\tau)=\xi)}{p(x(t)=\mu,y(t)=\xi,x(t+\tau)=\gamma,y(t+\tau)=\delta)}\right).
\end{eqnarray}

The BLRM is ergodic, therefore the densities $p(\zeta_{\tau}^{xy})$ and $p(\widetilde{\zeta_{\tau}^{xy}})$ can be empirically sampled looking at a single infinitely-long trajectory.

The causal structure of the BLRM (and of any signal-response model, see Fig.\ref{graph_SRM}) is such that the evolution of the signal is not influenced by the response, $p(x(t+\tau)|x(t),y(t))=p(x(t+\tau)|x(t))$. Then we can write the joint probability densities $p(\zeta_{\tau}^{xy})$ of couples of successive states over a time interval $\tau$ of the original trajectory as:

\begin{eqnarray}\label{A2}
&p(\zeta_{\tau}^{xy})\equiv p(x(t)=\gamma,y(t)=\delta,x(t+\tau)=\mu,y(t+\tau)=\xi)=\nonumber\\&=p(x(t)=\gamma)\cdot p(y(t)=\delta|x(t)=\gamma)\cdot p(x(t+\tau)=\mu|x(t)=\gamma)\cdot p(y(t+\tau)=\xi|x(t)=\gamma,y(t)=\delta,x(t+\tau)=\mu).~~~~
\end{eqnarray}

We need to evaluate all these probabilities. Since we are dealing with linear models, these are all Gaussian distributed, and we will calculate only the expected value and the variance of the relevant variables involved.

The system is Markovian, $p(x(t+\tau)|x(t+t'),x(t))=p(x(t+\tau)|x(t+t'))$ with $0\leq t'\leq \tau$, and the Bayes rule assumes the form $p(x(t+t')|x(t),x(t+\tau))=\frac{p(x(t+t')|x(t)) p(x(t+\tau)|x(t+t'))}{p(x(t+\tau)|x(t))}$. Then we calculate the conditional expected value for the signal $x(t+\tau)$ given a condition for its past $x(t)$ and another condition for its future $x(t+\tau)$ as: 

\begin{eqnarray}
\braket{x(t+t')|x(t),x(t+\tau)}=x(t)e^{-\frac{t'}{t_{rel}}}\frac{1-e^{-\frac{2(\tau-t')}{t_{rel}}}}{1-e^{-\frac{2\tau}{t_{rel}}}}+ x(t+\tau)e^{-\frac{\tau-t'}{t_{rel}}}\frac{1-e^{-\frac{2t'}{t_{rel}}}}{1-e^{-\frac{2\tau}{t_{rel}}}}.
\end{eqnarray}

Now we can calculate the full-conditional expectation of the response:

\begin{eqnarray}
&\braket{y(t+\tau)|x(t),y(t),x(t+\tau)}= y(t) e^{-\beta \tau}+\alpha\int_{0}^{\tau}dt'e^{-\beta(\tau-t')}\braket{x(t+t')|x(t),x(t+\tau)}=\nonumber\\&=y(t)e^{-\beta \tau}+\alpha \frac{e^{-\beta \tau}}{1-e^{-\frac{2\tau}{t_{rel}}}}\left(x(t)(\frac{e^{\tau (\beta-\frac{1}{t_{rel}})}-1}{\beta-\frac{1}{t_{rel}}}-\frac{e^{\tau(\beta-\frac{1}{t_{rel}})}-e^{-\frac{2\tau}{t_{rel}}}}{\beta+\frac{1}{t_{rel}}}) +x(t+\tau)(\frac{e^{\beta\tau }-e^{-\frac{\tau}{t_{rel}}}}{\beta+\frac{1}{t_{rel}}}-\frac{e^{\tau(\beta-\frac{2}{t_{rel}})}-e^{-\frac{\tau}{t_{rel}}}}{\beta-\frac{1}{t_{rel}}})\right).
\end{eqnarray}

One can immediately check that the limits for small and large time intervals $\tau$ verify respectively $\lim\limits_{\tau\rightarrow 0}\braket{y(t+\tau)|x(t),y(t),x(t+\tau)}=y(t)$ and $\lim\limits_{\tau\rightarrow \infty}\braket{y(t+\tau)|x(t),y(t),x(t+\tau)}=x(t+\tau)\frac{\alpha t_{rel}}{\beta t_{rel}+1}=\braket{y(t+\tau)|x(t+\tau)}$.

The causal order for the evolution of the signal is such that $p(x(t+t'')|x(t),x(t+t'),x(t+\tau))=p(x(t+t'')|x(t+t'),x(t+\tau))$ if $0\leq t'\leq t''\leq \tau$. Then we can calculate:

\begin{eqnarray}
&\braket{x(t+t')x(t+t'')|x(t),x(t+\tau)}_{t''\geq t'}=\nonumber\\&=\int_{-\infty}^{\infty}dx(t+t')p(x(t+t')|x(t),x(t+\tau))x(t+t')\braket{x(t+t'')|x(t+t'),x(t+\tau)}=\nonumber\\&=\braket{x(t+t')|x(t),x(t+\tau)}*\nonumber\\&*\left(x(t+\tau)e^{-\frac{\tau-t''}{t_{rel}}}\frac{\sigma^2_{t''-t'}}{\sigma^2_{\tau-t'}} +e^{-\frac{t''-t'}{t_{rel}}}\frac{\sigma^2_{\tau-t''}}{\sigma^2_{\tau-t'}}\left(\frac{1}{x(t)\frac{e^{-\frac{t'}{t_{rel}}}}{\sigma^2_{t'}}+x(t+\tau)\frac{e^{-\frac{\tau-t'}{t_{rel}}}}{\sigma^2_{\tau-t'}}}+\frac{x(t)\frac{e^{-\frac{t'}{t_{rel}}}}{\sigma^2_{t'}}+x(t+\tau)\frac{e^{-\frac{\tau-t'}{t_{rel}}}}{\sigma^2_{\tau-t'}}}{\frac{1}{\sigma^2_{t'}}+\frac{e^{-\frac{2(\tau-t')}{t_{rel}}}}{\sigma^2_{\tau-t'}}}\right)\right).
\end{eqnarray}

Let us write the full-conditional expectation of the squared response as a function of the expectations we just calculated:

\begin{eqnarray}
&\braket{y^2(t+\tau)|x(t),y(t),x(t+\tau)}=y^2(t) e^{-2\beta \tau}+2\alpha y(t) e^{-2\beta \tau} \int_{0}^{\tau}dt'e^{\beta t'}\braket{x(t+t')|x(t),x(t+\tau)}+ \nonumber\\& +\alpha^2e^{-2 \beta \tau}\int_{0}^{\tau}\int_{0}^{\tau}dt'dt''e^{\beta(t'+t'')}\braket{x(t+t')x(t+t'')|x(t),x(t+\tau)}.
\end{eqnarray}

A relevant feature of linear response models is that the conditional variances do not depend on the particular values of the conditioning variables\citep{auconi2017causal}. Here we consider the full-conditional variance $\sigma^2_{y(t+\tau)|x(t),y(t),x(t+\tau)}=\braket{y^2(t+\tau)|x(t),y(t),x(t+\tau)}-\braket{y(t+\tau)|x(t),y(t),x(t+\tau)}^2$, and it will be independent of the conditions $x(t)$, $y(t)$, and $x(t+\tau)$. Then the remaining terms in $\sigma^2_{y(t+\tau)|x(t),y(t),x(t+\tau)}$ sum up to:

\begin{eqnarray}
&\sigma^2_{y(t+\tau)|x(t),y(t),x(t+\tau)}=2\frac{\alpha^2 e^{-2 \beta \tau}}{\sigma^2_{\tau}} \int_{0}^{\tau}dt'' \sigma^2_{\tau-t''}e^{t''(\beta-\frac{1}{t_{rel}})}\int_{0}^{t''}dt' \sigma^2_{t'} e^{t'(\beta+\frac{1}{t_{rel}})}=2 \alpha^2 \sigma^2_x \frac{e^{-2\beta \tau}}{1-e^{-\frac{2\tau}{t_{rel}}}}*\\&* \left(-\frac{2}{t_{rel}}\frac{\beta+\frac{1}{t_{rel}}-\frac{2}{t_{rel}}e^{\tau(\beta-\frac{1}{t_{rel}})}-(\beta-\frac{1}{t_{rel}})e^{-\frac{2\tau}{t_{rel}}}}{(\beta+\frac{1}{t_{rel}})^2(\beta-\frac{1}{t_{rel}})^2} +\frac{\frac{1}{t_{rel}}e^{2\beta\tau}-\beta-\frac{1}{t_{rel}}+\beta e^{-\frac{2\tau}{t_{rel}}}}{2\beta(\beta+\frac{1}{t_{rel}})^2} -\frac{\frac{1}{t_{rel}}e^{2\tau(\beta-\frac{1}{t_{rel}})}-\beta+(\beta-\frac{1}{t_{rel}})e^{-\frac{2\tau}{t_{rel}}}}{2\beta(\beta-\frac{1}{t_{rel}})^2}\right),
\end{eqnarray}

where we used the fact that $\braket{x(t+t')x(t+t'')|x(t),x(t+\tau)}$ is symmetric in $t'$ and $t''$. We recall that for functions with the symmetry $f(t',t'')=f(t'',t')$ it holds: $\int_{0}^{\tau}\int_{0}^{\tau}dt' dt'' f(t',t'')=2\int_{0}^{\tau}dt'\int_{0}^{t'} dt'' f(t',t'')$.

The limits for small and large time intervals $\tau$ verify respectively $\lim\limits_{\tau\rightarrow 0}\sigma^2_{y(t+\tau)|x(t),y(t),x(t+\tau)}=0$ and $\lim\limits_{\tau\rightarrow \infty}\sigma^2_{y(t+\tau)|x(t),y(t),x(t+\tau)}=\alpha^2 \sigma^2_x \frac{t_{rel}}{\beta(\beta t_{rel}+1)^2}=\sigma^2_{y(t)|x(t)}$.

The factorization of the probability density $p(\zeta_\tau ^{xy})$ into conditional densities (Eq.\ref{A2}) leads to a decomposition of the mapping irreversibility. Here we show that in the BLRM all of these terms are zero except for the two terms corresponding to the full-conditional density of the evolution of the response in the original trajectory and in the time-reversed conjugate. 

For a stationary stochastic process like the BLRM it holds $p(x(t)=\gamma,y(t)=\delta)=p(x(t+\tau)=\gamma,y(t+\tau)=\delta)$, then these two terms cancel: 

\begin{eqnarray}
&\int_{-\infty}^{\infty}\int_{-\infty}^{\infty}\int_{-\infty}^{\infty}\int_{-\infty}^{\infty} d\gamma d\delta d\mu d\xi~ p(x(t)=\gamma,y(t)=\delta,x(t+\tau)=\mu,y(t+\tau)=\xi)\cdot \ln(p(x(t)=\gamma,y(t)=\delta))=\nonumber\\&=\int_{-\infty}^{\infty}\int_{-\infty}^{\infty} d\gamma d\delta~ p(x(t)=\gamma,y(t)=\delta)\cdot \ln(p(x(t)=\gamma,y(t)=\delta))=\nonumber\\&=\int_{-\infty}^{\infty}\int_{-\infty}^{\infty} d\gamma d\delta~ p(x(t+\tau)=\gamma,y(t+\tau)=\delta)\cdot \ln(p(x(t)=\gamma,y(t)=\delta))=\nonumber\\&=\int_{-\infty}^{\infty}\int_{-\infty}^{\infty} d\mu d\xi~ p(x(t+\tau)=\mu,y(t+\tau)=\xi)\cdot \ln(p(x(t)=\mu,y(t)=\xi))=\nonumber\\&=\int_{-\infty}^{\infty}\int_{-\infty}^{\infty}\int_{-\infty}^{\infty}\int_{-\infty}^{\infty} d\gamma d\delta d\mu d\xi~ p(x(t)=\gamma,y(t)=\delta,x(t+\tau)=\mu,y(t+\tau)=\xi)\cdot \ln(p(x(t)=\mu,y(t)=\xi)).
\end{eqnarray}

The contribution from the signal in the mapping irreversibility is also zero since the Ornstein-Uhlenbeck process is reversible, $p(x(t)=\gamma,x(t+\tau)=\mu)=p(x(t)=\mu,x(t+\tau)=\gamma)$:

\begin{eqnarray}
\int_{-\infty}^{\infty}\int_{-\infty}^{\infty} d\gamma d\mu ~ p(x(t)=\gamma,x(t+\tau)=\mu) ~ \ln\left(\frac{p(x(t+\tau)=\mu|x(t)=\gamma)}{p(x(t+\tau)=\gamma|x(t)=\mu)}\right)=0.
\end{eqnarray}

The mapping irreversibility is therefore:

\begin{eqnarray}
&\varPhi_\tau^{xy}=\braket{\varphi_\tau^{xy}}_{p(\zeta_\tau^{xy})}= \int_{-\infty}^{\infty}\int_{-\infty}^{\infty}\int_{-\infty}^{\infty}\int_{-\infty}^{\infty} d\gamma d\delta d\mu d\xi~ p(x(t)=\gamma,y(t)=\delta,x(t+\tau)=\mu,y(t+\tau)=\xi)*\nonumber\\ &*\ln\left(\frac{p(y(t+\tau)=\xi|x(t)=\gamma,y(t)=\delta,x(t+\tau)=\mu)}{p(y(t+\tau)=\delta|x(t)=\mu,y(t)=\xi,x(t+\tau)=\gamma)}\right)=\nonumber\\&=-\frac{1}{2}+\frac{1}{2\sigma^2_{y(t+\tau)|x(t),y(t),x(t+\tau)}} \int_{-\infty}^{\infty}\int_{-\infty}^{\infty}\int_{-\infty}^{\infty}\int_{-\infty}^{\infty} d\gamma d\delta d\mu d\xi~ p(x(t)=\gamma,y(t)=\delta,x(t+\tau)=\mu,y(t+\tau)=\xi)*\nonumber\\&*(\delta-\braket{y(t+\tau)|x(t)=\mu,y(t)=\xi,x(t+\tau)=\gamma})^2,
\end{eqnarray}

where in the last passage we exploited the fact that all the probability densities are Gaussian distributed. Solving the integrals we get the mapping irreversibility for the BLRM as a function of the time interval $\tau$:

\begin{eqnarray}
\varPhi_\tau^{xy}=\frac{1}{2}(e^{-2\beta\tau}-1)+\frac{2\alpha^2 \sigma^2_x t_{rel}^2}{\sigma^2_{y(t+\tau)|x(t),y(t),x(t+\tau)}} \frac{(e^{-\beta\tau}-e^{-\frac{\tau}{t_{rel}}})^2 (e^{-2\beta\tau}+1-2 e^{-\tau(\beta+\frac{1}{t_{rel}})})}{(\beta^2 t_{rel}^2-1)^2(1-e^{-\frac{2\tau}{t_{rel}}})}.
\end{eqnarray}

$\sigma^2_{y(t+\tau)|x(t),y(t),x(t+\tau)}$ is proportional to $\alpha^2 \sigma^2_x$, therefore the mapping irreversibility $\varPhi_\tau^{xy}$ is a function of just the two parameters $t_{rel}$ and $\beta$ (and of the observational time $\tau$).

\begin{figure}
	\begin{center}
		\includegraphics[scale=0.35]{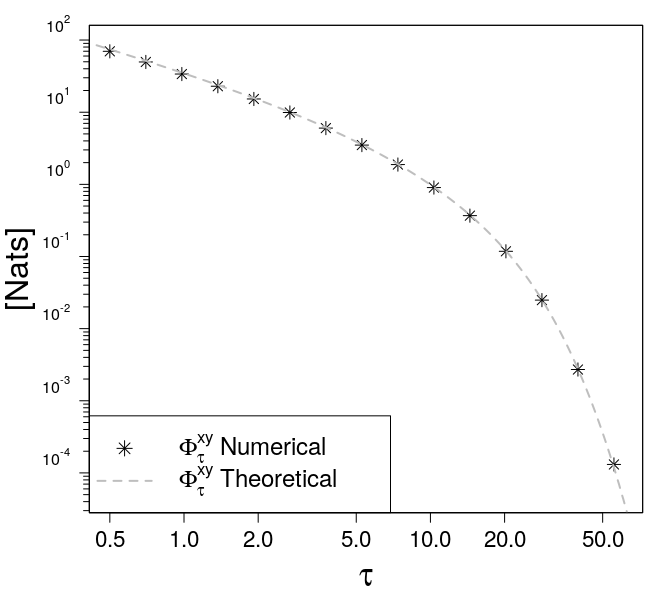}
		\caption{\label{Four} Numerical verification of the analytical solution for the entropy production $\varPhi_\tau^{xy}$ with observational time $\tau$ in the BLRM. The parameters are $\beta=0.2$ and $t_{rel}=10$.}
	\end{center}
\end{figure}

\section*{APPENDIX B: backward transfer ENTROPY IN THE BLRM}

In the BLRM, where all densities are Gaussian distributed, the backward transfer entropy is equivalent to the time-reversed Granger causality\citep{barnett2009granger}:

\begin{eqnarray}
T_{y\rightarrow x}(-\tau)=I(x(t),y(t+\tau)|x(t+\tau))=\ln\left(\frac{\sigma_{y|x}}{\sigma_{y(t+\tau)|x(t),x(t+\tau)}}\right)
\end{eqnarray}

We have to calculate the conditional variance $\sigma_{y(t+\tau)|x(t),x(t+\tau)}$. Let us recall the relation for the value of the response as a function of the whole past history of the signal trajectory:

\begin{eqnarray}
y(t+\tau)=\alpha e^{-\beta(t+\tau)} \left(\int_{-\infty} ^{t} dt' x(t') e^{\beta t'} +\int_{t} ^{t+\tau}dt' x(t') e^{\beta t'}\right)
\end{eqnarray}

Then we write the conditional squared response as

\begin{eqnarray}
&\braket{y^2(t+\tau)|x(t),x(t+\tau)}=2\alpha^2 e^{-2\beta (t+\tau)} ( e^{2\beta t}\int_{0}^{\tau} \int_{0}^{t''}dt'' dt' e^{\beta(t'+t'')}\braket{x(t+t')x(t+t'')|x(t),x(t+\tau)}_{t''\geq t'} +\nonumber\\& + \int_{-\infty}^{t} \int_{-\infty}^{t''} dt'' dt' \braket{x^2(t'')|x(t)} e^{-\frac{t''-t'}{t_{rel}}+\beta(t'+t'')} +\int_{-\infty}^{t} \int_{t}^{t+\tau} dt' dt'' \braket{x(t')|x(t)}\braket{x(t'')|x(t),x(t+\tau)} e^{\beta (t'+t'')} )  ~~
\end{eqnarray}

Since $\sigma^2_{y(t+\tau)|x(t),x(t+\tau)}$ is expected to be independent of $x(t)$ and $x(t+\tau)$, then the remaining terms sum up to:

\begin{eqnarray}
\sigma^2_{y(t+\tau)|x(t),x(t+\tau)}=\sigma^2_{y(t+\tau)|x(t),y(t),x(t+\tau)}+\sigma^2_{y|x} e^{-2\beta\tau},
\end{eqnarray}
where $\sigma^2_{y|x}=\frac{\sigma^2_x \alpha^2}{\beta t_{rel}(\beta+\frac{1}{t_{rel}})^2}$ was already calculated in \cite{auconi2017causal}. Then the backward transfer entropy is:

\begin{eqnarray}
T_{y\rightarrow x}(-\tau)=I\left( x(t),y(t)|x(t+\tau)\right)=-\frac{1}{2}\ln\left(\frac{\sigma^2_{y(t+\tau)|x(t),y(t),x(t+\tau)}}{\sigma^2_{y|x}}+e^{-2\beta\tau}\right).
\end{eqnarray}

\section*{APPENDIX C: LARGE OBSERVATIONAL TIME LIMIT OF THE ALTERNATIVE DEFINITION OF ENTROPY PRODUCTION IN THE BLRM.}

We are in the BLRM and the absence of feedback implies $p_\lambda(\zeta_\tau^{xy})=p(\zeta_\tau^{xy})$. Still the conditioning in the alternative definition (Eq.\ref{Ito-Sagawa}) is made on a different protocol compared to our definition of mapping irreversibility in controlled systems (Eq.\ref{detailed fluctuation theorem}). The effect of these different choices is best seen in the limit $\tau\rightarrow \infty$.

In the limit of large observational time, the probability density $p(\zeta_\tau^{xy})$, our definition $\varphi^{y|\lambda(x)}_\tau$ and the alternative definition $\eta^{y|\lambda(x)}_\tau $ decompose respectively into:

\begin{eqnarray}
&p(\zeta_\tau^{xy})=p(x(t)=x_t,y(t)=y_t,x(t+\tau)=x_{t+\tau},y(t+\tau)=y_{t+\tau})\nonumber \\&\rightarrow p(x(t)=x_t,y(t)=y_t)\cdot p(x(t+\tau)=x_{t+\tau},y(t+\tau)=y_{t+\tau}),
\end{eqnarray}

\begin{eqnarray}
\varphi^{y|\lambda(x)}_\tau \rightarrow \ln\left(\frac{p(y_{t+\tau}|x_{t+\tau})}{p(\widetilde{y_{t+\tau}}|\widetilde{x_{t+\tau}})}\right)=\ln\left(\frac{p(y_{t+\tau}|x_{t+\tau})}{p(y_t|x_t)}\right),
\end{eqnarray}

\begin{eqnarray}
\eta^{y|\lambda(x)}_\tau \rightarrow \ln\left(\frac{p(y_{t+\tau}|x_{t+\tau})}{p(\widetilde{y_{t+\tau}}|x_{t+\tau})}\right),
\end{eqnarray}

where we neglected the change of entropy in the $y$ system, that is $\ln\left(\frac{p(y_t)}{p(y_{t+\tau})}\right)$, because its ensemble average vanishes in the BLRM.

$\left\langle\varphi^{y|\lambda(x)}_\tau\right\rangle_{p(\zeta_\tau^{xy})}$ is vanishing in stationary systems in the limit $\tau\rightarrow \infty$ as a consequence of the factorization of $p(\zeta_\tau^{xy})$.

The ensemble average of the alternative definition is not vanishing and it is calculated as:

\begin{eqnarray}
&\braket{\eta^{y|x}_\tau}=-\frac{1}{2}\left(1+\ln(2\pi\sigma^2_{y|x})\right)+\frac{1}{2}\int_{-\infty}^{\infty}\int_{-\infty}^{\infty}dx_{t+\tau}dy_t p(x_{t+\tau})p(y_t) \left(\ln(2\pi\sigma^2_{y|x})+\frac{(\widetilde{y_{t+\tau}}-\braket{\widetilde{y_{t+\tau}}|x_{t+\tau}})^2}{\sigma^2_{y|x}}\right)=\nonumber\\&=-\frac{1}{2}+\frac{1}{2}\frac{\sigma^2_y}{\sigma^2_{y|x}}\left(1+\frac{\beta t_{rel}}{\beta t_{rel}+1}\right)=\beta t_{rel},
\end{eqnarray}

where we used the relations $p(\widetilde{y_{t+\tau}}|x_{t+\tau})=p(y(t+\tau)=y_t|x(t+\tau)=x_{t+\tau})$, $\braket{\widetilde{y_{t+\tau}}|x_{t+\tau}}=\braket{y_{t+\tau}|x_{t+\tau}}=\braket{y(t+\tau)|x(t+\tau)=x_{t+\tau}}=\braket{y(t)|x(t)=x_{t+\tau}}=x_{t+\tau} \frac{\alpha t_{rel}}{\beta t_{rel}+1}$, and $\sigma^2_y=(1+\beta t_{rel})\sigma^2_{y|x}=\sigma^2_x \frac{\alpha^2 t_{rel}}{\beta(\beta t_{rel}+1)}$ (such quantities and also the mutual information and standard transfer entropy in the BLRM are discussed in \citep{auconi2017causal}).

\section*{APPENDIX D: NUMERICAL ESTIMATION OF THE ENTROPY PRODUCTION IN THE BIVARIATE GAUSSIAN APPROXIMATION.}

We calculate numerically the mapping irreversibility $\varPhi_\tau^{xy}$ as an average of the spatial density of entropy production $\psi(x_t,y_t)$, $\varPhi_\tau^{xy}=\int^\infty _{-\infty}\int^\infty _{-\infty} dx_t dy_t \psi(x_t,y_t)$. In the computer algorithm the $(x,y)$ space is dicretized in boxes $(i,j)$, and for each box we estimate the conditional correlation $C_{xy|i,j}$ of future values $(x_{t+\tau},y_{t+\tau})$, the conditional correlation $\widetilde{C_{xy|i,j}}$ of past values $(x_{t-\tau},y_{t-\tau})$, the expected values for both variables in future ($\braket{x|i,j}$, $\braket{y|i,j}$) and past states ($\braket{\widetilde{x}|i,j}$, $\braket{\widetilde{y}|i,j}$), and the standard deviations on those $\sigma_{x|i,j}$, $\widetilde{\sigma_{x|i,j}}$, $\sigma_{y|i,j}$, $\widetilde{\sigma_{y|i,j}}$. The spacial density evaluated in the box $(i,j)$ is then calculated as the bidimensional Kullback-Leibler divergence in the Gaussian approximation\cite{duchi2007derivations}:

\begin{eqnarray}
& \psi(i,j) = P(i,j) ~ \frac{1}{2} [ \ln(\frac{\widetilde{\sigma_{x|i,j}^2}\widetilde{\sigma_{y|i,j}^2}(1-\widetilde{C^2_{xy|i,j}})}{\sigma_{x|i,j}^2\sigma_{y|i,j}^2(1-C^2_{xy|i,j})})  -2 + \frac{\frac{\sigma_{x|i,j}^2}{\widetilde{\sigma_{x|i,j}^2}}+\frac{\sigma_{y|i,j}^2}{\widetilde{\sigma_{y|i,j}^2}}-2\frac{\sigma_{x|i,j}\sigma_{y|i,j}}{\widetilde{\sigma_{x|i,j}}\widetilde{\sigma_{y|i,j}}}C_{xy|i,j}\widetilde{C_{xy|i,j}}}{1-\widetilde{C^2_{xy|i,j}}}	+\nonumber\\& +\frac{\widetilde{\sigma^2_{y|i,j}}(\braket{\widetilde{x}|i,j}-\braket{x|i,j})^2 + \widetilde{\sigma^2_{x|i,j}}(\braket{\widetilde{y}|i,j}-\braket{y|i,j})^2 -2 \widetilde{C_{xy|i,j}}\widetilde{\sigma_{x|i,j}}\widetilde{\sigma_{y|i,j}}(\braket{\widetilde{x}|i,j}-\braket{x|i,j})(\braket{\widetilde{y}|i,j}-\braket{y|i,j})}{\widetilde{\sigma^2_{x|i,j}}\widetilde{\sigma^2_{y|i,j}}(1-\widetilde{C^2_{xy|i,j}}) } ]
\end{eqnarray}

The effect of the finite width of the discretization is attenuated by estimating all the quantities taking into account the starting point $(x_t,y_t)$ within the box $(i,j)$, subtracting the difference to the mean values for each box. For example, when we sample for the estimate of the conditional average $\braket{x_{t+\tau}|i}$ we would collect samples $x_{t+\tau}-(x_t-\braket{x_t|i})$.

\end{widetext}

\bibliography{Paper}

\providecommand{\noopsort}[1]{}\providecommand{\singleletter}[1]{#1}%
\begin{thebibliography}{36}%
\makeatletter
\providecommand \@ifxundefined [1]{%
 \@ifx{#1\undefined}
}%
\providecommand \@ifnum [1]{%
 \ifnum #1\expandafter \@firstoftwo
 \else \expandafter \@secondoftwo
 \fi
}%
\providecommand \@ifx [1]{%
 \ifx #1\expandafter \@firstoftwo
 \else \expandafter \@secondoftwo
 \fi
}%
\providecommand \natexlab [1]{#1}%
\providecommand \enquote  [1]{``#1''}%
\providecommand \bibnamefont  [1]{#1}%
\providecommand \bibfnamefont [1]{#1}%
\providecommand \citenamefont [1]{#1}%
\providecommand \href@noop [0]{\@secondoftwo}%
\providecommand \href [0]{\begingroup \@sanitize@url \@href}%
\providecommand \@href[1]{\@@startlink{#1}\@@href}%
\providecommand \@@href[1]{\endgroup#1\@@endlink}%
\providecommand \@sanitize@url [0]{\catcode `\\12\catcode `\$12\catcode
  `\&12\catcode `\#12\catcode `\^12\catcode `\_12\catcode `\%12\relax}%
\providecommand \@@startlink[1]{}%
\providecommand \@@endlink[0]{}%
\providecommand \url  [0]{\begingroup\@sanitize@url \@url }%
\providecommand \@url [1]{\endgroup\@href {#1}{\urlprefix }}%
\providecommand \urlprefix  [0]{URL }%
\providecommand \Eprint [0]{\href }%
\providecommand \doibase [0]{http://dx.doi.org/}%
\providecommand \selectlanguage [0]{\@gobble}%
\providecommand \bibinfo  [0]{\@secondoftwo}%
\providecommand \bibfield  [0]{\@secondoftwo}%
\providecommand \translation [1]{[#1]}%
\providecommand \BibitemOpen [0]{}%
\providecommand \bibitemStop [0]{}%
\providecommand \bibitemNoStop [0]{.\EOS\space}%
\providecommand \EOS [0]{\spacefactor3000\relax}%
\providecommand \BibitemShut  [1]{\csname bibitem#1\endcsname}%
\let\auto@bib@innerbib\@empty
\bibitem [{\citenamefont {Jarzynski}(2011)}]{jarzynski2011equalities}%
  \BibitemOpen
  \bibfield  {author} {\bibinfo {author} {\bibfnamefont {C.}~\bibnamefont
  {Jarzynski}},\ }\href@noop {} {\bibfield  {journal} {\bibinfo  {journal}
  {Annu. Rev. Condens. Matter Phys.}\ }\textbf {\bibinfo {volume} {2}},\
  \bibinfo {pages} {329} (\bibinfo {year} {2011})}\BibitemShut {NoStop}%
\bibitem [{\citenamefont {Parrondo}\ \emph {et~al.}(2009)\citenamefont
  {Parrondo}, \citenamefont {Van~den Broeck},\ and\ \citenamefont
  {Kawai}}]{parrondo2009entropy}%
  \BibitemOpen
  \bibfield  {author} {\bibinfo {author} {\bibfnamefont {J.~M.}\ \bibnamefont
  {Parrondo}}, \bibinfo {author} {\bibfnamefont {C.}~\bibnamefont {Van~den
  Broeck}}, \ and\ \bibinfo {author} {\bibfnamefont {R.}~\bibnamefont
  {Kawai}},\ }\href@noop {} {\bibfield  {journal} {\bibinfo  {journal} {New
  Journal of Physics}\ }\textbf {\bibinfo {volume} {11}},\ \bibinfo {pages}
  {073008} (\bibinfo {year} {2009})}\BibitemShut {NoStop}%
\bibitem [{\citenamefont {Feng}\ and\ \citenamefont
  {Crooks}(2008)}]{feng2008length}%
  \BibitemOpen
  \bibfield  {author} {\bibinfo {author} {\bibfnamefont {E.~H.}\ \bibnamefont
  {Feng}}\ and\ \bibinfo {author} {\bibfnamefont {G.~E.}\ \bibnamefont
  {Crooks}},\ }\href@noop {} {\bibfield  {journal} {\bibinfo  {journal}
  {Physical review letters}\ }\textbf {\bibinfo {volume} {101}},\ \bibinfo
  {pages} {090602} (\bibinfo {year} {2008})}\BibitemShut {NoStop}%
\bibitem [{\citenamefont {Jarzynski}(1997)}]{jarzynski1997nonequilibrium}%
  \BibitemOpen
  \bibfield  {author} {\bibinfo {author} {\bibfnamefont {C.}~\bibnamefont
  {Jarzynski}},\ }\href@noop {} {\bibfield  {journal} {\bibinfo  {journal}
  {Physical Review Letters}\ }\textbf {\bibinfo {volume} {78}},\ \bibinfo
  {pages} {2690} (\bibinfo {year} {1997})}\BibitemShut {NoStop}%
\bibitem [{\citenamefont {Crooks}(1999)}]{crooks1999entropy}%
  \BibitemOpen
  \bibfield  {author} {\bibinfo {author} {\bibfnamefont {G.~E.}\ \bibnamefont
  {Crooks}},\ }\href@noop {} {\bibfield  {journal} {\bibinfo  {journal}
  {Physical Review E}\ }\textbf {\bibinfo {volume} {60}},\ \bibinfo {pages}
  {2721} (\bibinfo {year} {1999})}\BibitemShut {NoStop}%
\bibitem [{\citenamefont {Evans}\ and\ \citenamefont
  {Searles}(2002)}]{evans2002fluctuation}%
  \BibitemOpen
  \bibfield  {author} {\bibinfo {author} {\bibfnamefont {D.~J.}\ \bibnamefont
  {Evans}}\ and\ \bibinfo {author} {\bibfnamefont {D.~J.}\ \bibnamefont
  {Searles}},\ }\href@noop {} {\bibfield  {journal} {\bibinfo  {journal}
  {Advances in Physics}\ }\textbf {\bibinfo {volume} {51}},\ \bibinfo {pages}
  {1529} (\bibinfo {year} {2002})}\BibitemShut {NoStop}%
\bibitem [{\citenamefont {Kawai}\ \emph {et~al.}(2007)\citenamefont {Kawai},
  \citenamefont {Parrondo},\ and\ \citenamefont {Van~den
  Broeck}}]{kawai2007dissipation}%
  \BibitemOpen
  \bibfield  {author} {\bibinfo {author} {\bibfnamefont {R.}~\bibnamefont
  {Kawai}}, \bibinfo {author} {\bibfnamefont {J.}~\bibnamefont {Parrondo}}, \
  and\ \bibinfo {author} {\bibfnamefont {C.}~\bibnamefont {Van~den Broeck}},\
  }\href@noop {} {\bibfield  {journal} {\bibinfo  {journal} {Physical review
  letters}\ }\textbf {\bibinfo {volume} {98}},\ \bibinfo {pages} {080602}
  (\bibinfo {year} {2007})}\BibitemShut {NoStop}%
\bibitem [{\citenamefont {Jarzynski}(2000)}]{jarzynski2000hamiltonian}%
  \BibitemOpen
  \bibfield  {author} {\bibinfo {author} {\bibfnamefont {C.}~\bibnamefont
  {Jarzynski}},\ }\href@noop {} {\bibfield  {journal} {\bibinfo  {journal}
  {Journal of Statistical Physics}\ }\textbf {\bibinfo {volume} {98}},\
  \bibinfo {pages} {77} (\bibinfo {year} {2000})}\BibitemShut {NoStop}%
\bibitem [{\citenamefont {Chernyak}\ \emph {et~al.}(2006)\citenamefont
  {Chernyak}, \citenamefont {Chertkov},\ and\ \citenamefont
  {Jarzynski}}]{chernyak2006path}%
  \BibitemOpen
  \bibfield  {author} {\bibinfo {author} {\bibfnamefont {V.~Y.}\ \bibnamefont
  {Chernyak}}, \bibinfo {author} {\bibfnamefont {M.}~\bibnamefont {Chertkov}},
  \ and\ \bibinfo {author} {\bibfnamefont {C.}~\bibnamefont {Jarzynski}},\
  }\href@noop {} {\bibfield  {journal} {\bibinfo  {journal} {Journal of
  Statistical Mechanics: Theory and Experiment}\ }\textbf {\bibinfo {volume}
  {2006}},\ \bibinfo {pages} {P08001} (\bibinfo {year} {2006})}\BibitemShut
  {NoStop}%
\bibitem [{\citenamefont {Ito}\ and\ \citenamefont
  {Sagawa}(2013)}]{ito2013information}%
  \BibitemOpen
  \bibfield  {author} {\bibinfo {author} {\bibfnamefont {S.}~\bibnamefont
  {Ito}}\ and\ \bibinfo {author} {\bibfnamefont {T.}~\bibnamefont {Sagawa}},\
  }\href@noop {} {\bibfield  {journal} {\bibinfo  {journal} {Physical review
  letters}\ }\textbf {\bibinfo {volume} {111}},\ \bibinfo {pages} {180603}
  (\bibinfo {year} {2013})}\BibitemShut {NoStop}%
\bibitem [{\citenamefont {Sagawa}\ and\ \citenamefont
  {Ueda}(2012)}]{sagawa2012nonequilibrium}%
  \BibitemOpen
  \bibfield  {author} {\bibinfo {author} {\bibfnamefont {T.}~\bibnamefont
  {Sagawa}}\ and\ \bibinfo {author} {\bibfnamefont {M.}~\bibnamefont {Ueda}},\
  }\href@noop {} {\bibfield  {journal} {\bibinfo  {journal} {Physical Review
  E}\ }\textbf {\bibinfo {volume} {85}},\ \bibinfo {pages} {021104} (\bibinfo
  {year} {2012})}\BibitemShut {NoStop}%
\bibitem [{\citenamefont {Sagawa}\ and\ \citenamefont
  {Ueda}(2010)}]{sagawa2010generalized}%
  \BibitemOpen
  \bibfield  {author} {\bibinfo {author} {\bibfnamefont {T.}~\bibnamefont
  {Sagawa}}\ and\ \bibinfo {author} {\bibfnamefont {M.}~\bibnamefont {Ueda}},\
  }\href@noop {} {\bibfield  {journal} {\bibinfo  {journal} {Physical review
  letters}\ }\textbf {\bibinfo {volume} {104}},\ \bibinfo {pages} {090602}
  (\bibinfo {year} {2010})}\BibitemShut {NoStop}%
\bibitem [{\citenamefont {Szilard}(1964)}]{szilard1964decrease}%
  \BibitemOpen
  \bibfield  {author} {\bibinfo {author} {\bibfnamefont {L.}~\bibnamefont
  {Szilard}},\ }\href@noop {} {\bibfield  {journal} {\bibinfo  {journal}
  {Systems Research and Behavioral Science}\ }\textbf {\bibinfo {volume} {9}},\
  \bibinfo {pages} {301} (\bibinfo {year} {1964})}\BibitemShut {NoStop}%
\bibitem [{\citenamefont {Toyabe}\ \emph {et~al.}(2010)\citenamefont {Toyabe},
  \citenamefont {Sagawa}, \citenamefont {Ueda}, \citenamefont {Muneyuki},\ and\
  \citenamefont {Sano}}]{toyabe2010experimental}%
  \BibitemOpen
  \bibfield  {author} {\bibinfo {author} {\bibfnamefont {S.}~\bibnamefont
  {Toyabe}}, \bibinfo {author} {\bibfnamefont {T.}~\bibnamefont {Sagawa}},
  \bibinfo {author} {\bibfnamefont {M.}~\bibnamefont {Ueda}}, \bibinfo {author}
  {\bibfnamefont {E.}~\bibnamefont {Muneyuki}}, \ and\ \bibinfo {author}
  {\bibfnamefont {M.}~\bibnamefont {Sano}},\ }\href@noop {} {\bibfield
  {journal} {\bibinfo  {journal} {Nature Physics}\ }\textbf {\bibinfo {volume}
  {6}},\ \bibinfo {pages} {988} (\bibinfo {year} {2010})}\BibitemShut {NoStop}%
\bibitem [{\citenamefont {Auconi}\ \emph {et~al.}(2017)\citenamefont {Auconi},
  \citenamefont {Giansanti},\ and\ \citenamefont {Klipp}}]{auconi2017causal}%
  \BibitemOpen
  \bibfield  {author} {\bibinfo {author} {\bibfnamefont {A.}~\bibnamefont
  {Auconi}}, \bibinfo {author} {\bibfnamefont {A.}~\bibnamefont {Giansanti}}, \
  and\ \bibinfo {author} {\bibfnamefont {E.}~\bibnamefont {Klipp}},\
  }\href@noop {} {\bibfield  {journal} {\bibinfo  {journal} {Physical Review
  E}\ }\textbf {\bibinfo {volume} {95}},\ \bibinfo {pages} {042315} (\bibinfo
  {year} {2017})}\BibitemShut {NoStop}%
\bibitem [{\citenamefont {Cover}\ and\ \citenamefont
  {Thomas}(2012)}]{cover2012elements}%
  \BibitemOpen
  \bibfield  {author} {\bibinfo {author} {\bibfnamefont {T.~M.}\ \bibnamefont
  {Cover}}\ and\ \bibinfo {author} {\bibfnamefont {J.~A.}\ \bibnamefont
  {Thomas}},\ }\href@noop {} {\emph {\bibinfo {title} {Elements of information
  theory}}}\ (\bibinfo  {publisher} {John Wiley \& Sons},\ \bibinfo {year}
  {2012})\BibitemShut {NoStop}%
\bibitem [{\citenamefont {Ito}(2016)}]{ito2016backward}%
  \BibitemOpen
  \bibfield  {author} {\bibinfo {author} {\bibfnamefont {S.}~\bibnamefont
  {Ito}},\ }\href@noop {} {\bibfield  {journal} {\bibinfo  {journal}
  {Scientific reports}\ }\textbf {\bibinfo {volume} {6}} (\bibinfo {year}
  {2016})}\BibitemShut {NoStop}%
\bibitem [{\citenamefont {Seifert}(2012)}]{seifert2012stochastic}%
  \BibitemOpen
  \bibfield  {author} {\bibinfo {author} {\bibfnamefont {U.}~\bibnamefont
  {Seifert}},\ }\href@noop {} {\bibfield  {journal} {\bibinfo  {journal}
  {Reports on Progress in Physics}\ }\textbf {\bibinfo {volume} {75}},\
  \bibinfo {pages} {126001} (\bibinfo {year} {2012})}\BibitemShut {NoStop}%
\bibitem [{\citenamefont {Shreve}(2004)}]{shreve2004stochastic}%
  \BibitemOpen
  \bibfield  {author} {\bibinfo {author} {\bibfnamefont {S.~E.}\ \bibnamefont
  {Shreve}},\ }\href@noop {} {\emph {\bibinfo {title} {Stochastic calculus for
  finance II: Continuous-time models}}},\ Vol.~\bibinfo {volume} {11}\
  (\bibinfo  {publisher} {Springer Science \& Business Media},\ \bibinfo {year}
  {2004})\BibitemShut {NoStop}%
\bibitem [{\citenamefont {Seifert}(2005)}]{seifert2005entropy}%
  \BibitemOpen
  \bibfield  {author} {\bibinfo {author} {\bibfnamefont {U.}~\bibnamefont
  {Seifert}},\ }\href@noop {} {\bibfield  {journal} {\bibinfo  {journal}
  {Physical review letters}\ }\textbf {\bibinfo {volume} {95}},\ \bibinfo
  {pages} {040602} (\bibinfo {year} {2005})}\BibitemShut {NoStop}%
\bibitem [{\citenamefont {Ito}\ and\ \citenamefont
  {Sagawa}(2016)}]{ito2016information}%
  \BibitemOpen
  \bibfield  {author} {\bibinfo {author} {\bibfnamefont {S.}~\bibnamefont
  {Ito}}\ and\ \bibinfo {author} {\bibfnamefont {T.}~\bibnamefont {Sagawa}},\
  }\href@noop {} {\bibfield  {journal} {\bibinfo  {journal} {Mathematical
  Foundations and Applications of Graph Entropy}\ }\textbf {\bibinfo {volume}
  {3}},\ \bibinfo {pages} {63} (\bibinfo {year} {2016})}\BibitemShut {NoStop}%
\bibitem [{\citenamefont {Uhlenbeck}\ and\ \citenamefont
  {Ornstein}(1930)}]{uhlenbeck1930theory}%
  \BibitemOpen
  \bibfield  {author} {\bibinfo {author} {\bibfnamefont {G.~E.}\ \bibnamefont
  {Uhlenbeck}}\ and\ \bibinfo {author} {\bibfnamefont {L.~S.}\ \bibnamefont
  {Ornstein}},\ }\href@noop {} {\bibfield  {journal} {\bibinfo  {journal}
  {Physical review}\ }\textbf {\bibinfo {volume} {36}},\ \bibinfo {pages} {823}
  (\bibinfo {year} {1930})}\BibitemShut {NoStop}%
\bibitem [{\citenamefont {Gillespie}(1996)}]{gillespie1996exact}%
  \BibitemOpen
  \bibfield  {author} {\bibinfo {author} {\bibfnamefont {D.~T.}\ \bibnamefont
  {Gillespie}},\ }\href@noop {} {\bibfield  {journal} {\bibinfo  {journal}
  {Physical review E}\ }\textbf {\bibinfo {volume} {54}},\ \bibinfo {pages}
  {2084} (\bibinfo {year} {1996})}\BibitemShut {NoStop}%
\bibitem [{\citenamefont {{R Core Team}}(2014)}]{team2014r}%
  \BibitemOpen
  \bibfield  {author} {\bibinfo {author} {\bibnamefont {{R Core Team}}},\
  }\href@noop {} {\emph {\bibinfo {title} {R: A Language and Environment for
  Statistical Computing}}},\ \bibinfo {organization} {R Foundation for
  Statistical Computing},\ \bibinfo {address} {Vienna, Austria} (\bibinfo
  {year} {2014})\BibitemShut {NoStop}%
\bibitem [{\citenamefont {Klipp}\ \emph {et~al.}(2016)\citenamefont {Klipp},
  \citenamefont {Liebermeister}, \citenamefont {Wierling}, \citenamefont
  {Kowald},\ and\ \citenamefont {Herwig}}]{klipp2016systems}%
  \BibitemOpen
  \bibfield  {author} {\bibinfo {author} {\bibfnamefont {E.}~\bibnamefont
  {Klipp}}, \bibinfo {author} {\bibfnamefont {W.}~\bibnamefont
  {Liebermeister}}, \bibinfo {author} {\bibfnamefont {C.}~\bibnamefont
  {Wierling}}, \bibinfo {author} {\bibfnamefont {A.}~\bibnamefont {Kowald}}, \
  and\ \bibinfo {author} {\bibfnamefont {R.}~\bibnamefont {Herwig}},\
  }\href@noop {} {\emph {\bibinfo {title} {Systems biology: a textbook}}}\
  (\bibinfo  {publisher} {John Wiley \& Sons},\ \bibinfo {year}
  {2016})\BibitemShut {NoStop}%
\bibitem [{\citenamefont {Kholodenko}(2006)}]{kholodenko2006cell}%
  \BibitemOpen
  \bibfield  {author} {\bibinfo {author} {\bibfnamefont {B.~N.}\ \bibnamefont
  {Kholodenko}},\ }\href@noop {} {\bibfield  {journal} {\bibinfo  {journal}
  {Nature reviews Molecular cell biology}\ }\textbf {\bibinfo {volume} {7}},\
  \bibinfo {pages} {165} (\bibinfo {year} {2006})}\BibitemShut {NoStop}%
\bibitem [{\citenamefont {Bialek}\ and\ \citenamefont
  {Setayeshgar}(2005)}]{bialek2005physical}%
  \BibitemOpen
  \bibfield  {author} {\bibinfo {author} {\bibfnamefont {W.}~\bibnamefont
  {Bialek}}\ and\ \bibinfo {author} {\bibfnamefont {S.}~\bibnamefont
  {Setayeshgar}},\ }\href@noop {} {\bibfield  {journal} {\bibinfo  {journal}
  {Proceedings of the National Academy of Sciences of the United States of
  America}\ }\textbf {\bibinfo {volume} {102}},\ \bibinfo {pages} {10040}
  (\bibinfo {year} {2005})}\BibitemShut {NoStop}%
\bibitem [{\citenamefont {Tka{\v{c}}ik}\ \emph {et~al.}(2008)\citenamefont
  {Tka{\v{c}}ik}, \citenamefont {Callan},\ and\ \citenamefont
  {Bialek}}]{tkavcik2008information}%
  \BibitemOpen
  \bibfield  {author} {\bibinfo {author} {\bibfnamefont {G.}~\bibnamefont
  {Tka{\v{c}}ik}}, \bibinfo {author} {\bibfnamefont {C.~G.}\ \bibnamefont
  {Callan}}, \ and\ \bibinfo {author} {\bibfnamefont {W.}~\bibnamefont
  {Bialek}},\ }\href@noop {} {\bibfield  {journal} {\bibinfo  {journal}
  {Proceedings of the National Academy of Sciences}\ }\textbf {\bibinfo
  {volume} {105}},\ \bibinfo {pages} {12265} (\bibinfo {year}
  {2008})}\BibitemShut {NoStop}%
\bibitem [{\citenamefont {Crisanti}\ \emph {et~al.}(2018)\citenamefont
  {Crisanti}, \citenamefont {De~Martino},\ and\ \citenamefont
  {Fiorentino}}]{crisanti2018statistics}%
  \BibitemOpen
  \bibfield  {author} {\bibinfo {author} {\bibfnamefont {A.}~\bibnamefont
  {Crisanti}}, \bibinfo {author} {\bibfnamefont {A.}~\bibnamefont
  {De~Martino}}, \ and\ \bibinfo {author} {\bibfnamefont {J.}~\bibnamefont
  {Fiorentino}},\ }\href@noop {} {\bibfield  {journal} {\bibinfo  {journal}
  {Physical Review E}\ }\textbf {\bibinfo {volume} {97}},\ \bibinfo {pages}
  {022407} (\bibinfo {year} {2018})}\BibitemShut {NoStop}%
\bibitem [{\citenamefont {Waltermann}\ and\ \citenamefont
  {Klipp}(2011)}]{waltermann2011information}%
  \BibitemOpen
  \bibfield  {author} {\bibinfo {author} {\bibfnamefont {C.}~\bibnamefont
  {Waltermann}}\ and\ \bibinfo {author} {\bibfnamefont {E.}~\bibnamefont
  {Klipp}},\ }\href@noop {} {\bibfield  {journal} {\bibinfo  {journal}
  {Biochimica et Biophysica Acta (BBA)-General Subjects}\ }\textbf {\bibinfo
  {volume} {1810}},\ \bibinfo {pages} {924} (\bibinfo {year}
  {2011})}\BibitemShut {NoStop}%
\bibitem [{\citenamefont {Di~Talia}\ and\ \citenamefont
  {Wieschaus}(2012)}]{di2012short}%
  \BibitemOpen
  \bibfield  {author} {\bibinfo {author} {\bibfnamefont {S.}~\bibnamefont
  {Di~Talia}}\ and\ \bibinfo {author} {\bibfnamefont {E.~F.}\ \bibnamefont
  {Wieschaus}},\ }\href@noop {} {\bibfield  {journal} {\bibinfo  {journal}
  {Developmental cell}\ }\textbf {\bibinfo {volume} {22}},\ \bibinfo {pages}
  {763} (\bibinfo {year} {2012})}\BibitemShut {NoStop}%
\bibitem [{\citenamefont {Nemenman}(2012)}]{nemenman2012gain}%
  \BibitemOpen
  \bibfield  {author} {\bibinfo {author} {\bibfnamefont {I.}~\bibnamefont
  {Nemenman}},\ }\href@noop {} {\bibfield  {journal} {\bibinfo  {journal}
  {Physical biology}\ }\textbf {\bibinfo {volume} {9}},\ \bibinfo {pages}
  {026003} (\bibinfo {year} {2012})}\BibitemShut {NoStop}%
\bibitem [{\citenamefont {Di~Talia}\ and\ \citenamefont
  {Wieschaus}(2014)}]{di2014simple}%
  \BibitemOpen
  \bibfield  {author} {\bibinfo {author} {\bibfnamefont {S.}~\bibnamefont
  {Di~Talia}}\ and\ \bibinfo {author} {\bibfnamefont {E.~F.}\ \bibnamefont
  {Wieschaus}},\ }\href@noop {} {\bibfield  {journal} {\bibinfo  {journal}
  {Biophysical journal}\ }\textbf {\bibinfo {volume} {107}},\ \bibinfo {pages}
  {L1} (\bibinfo {year} {2014})}\BibitemShut {NoStop}%
\bibitem [{\citenamefont {Tka{\v{c}}ik}\ \emph {et~al.}(2009)\citenamefont
  {Tka{\v{c}}ik}, \citenamefont {Walczak},\ and\ \citenamefont
  {Bialek}}]{tkavcik2009optimizing}%
  \BibitemOpen
  \bibfield  {author} {\bibinfo {author} {\bibfnamefont {G.}~\bibnamefont
  {Tka{\v{c}}ik}}, \bibinfo {author} {\bibfnamefont {A.~M.}\ \bibnamefont
  {Walczak}}, \ and\ \bibinfo {author} {\bibfnamefont {W.}~\bibnamefont
  {Bialek}},\ }\href@noop {} {\bibfield  {journal} {\bibinfo  {journal}
  {Physical Review E}\ }\textbf {\bibinfo {volume} {80}},\ \bibinfo {pages}
  {031920} (\bibinfo {year} {2009})}\BibitemShut {NoStop}%
\bibitem [{\citenamefont {Barnett}\ \emph {et~al.}(2009)\citenamefont
  {Barnett}, \citenamefont {Barrett},\ and\ \citenamefont
  {Seth}}]{barnett2009granger}%
  \BibitemOpen
  \bibfield  {author} {\bibinfo {author} {\bibfnamefont {L.}~\bibnamefont
  {Barnett}}, \bibinfo {author} {\bibfnamefont {A.~B.}\ \bibnamefont
  {Barrett}}, \ and\ \bibinfo {author} {\bibfnamefont {A.~K.}\ \bibnamefont
  {Seth}},\ }\href@noop {} {\bibfield  {journal} {\bibinfo  {journal} {Physical
  review letters}\ }\textbf {\bibinfo {volume} {103}},\ \bibinfo {pages}
  {238701} (\bibinfo {year} {2009})}\BibitemShut {NoStop}%
\bibitem [{\citenamefont {Duchi}(2007)}]{duchi2007derivations}%
  \BibitemOpen
  \bibfield  {author} {\bibinfo {author} {\bibfnamefont {J.}~\bibnamefont
  {Duchi}},\ }\href@noop {} {\bibfield  {journal} {\bibinfo  {journal}
  {Berkeley, California}\ } (\bibinfo {year} {2007})}\BibitemShut {NoStop}%
\end{thebibliography}%

\end{document}